# Crystal structures of two furazidin polymorphs revealed by a joint effort of crystal structure prediction and NMR crystallography

Marta K. Dudek,[1]* Piotr Paluch,[1] Edyta Pindelska[2]*

[1]Centre of Molecular and Macromolecular Studies, Polish Academy of Sciences, Sienkiewicza 112, 90363 Lodz, Poland, e-mail: mdudek@cbmm.lodz.pl

[2]Faculty of Pharmacy, Medical University of Warsaw, Banacha 1, 02097 Warsaw, Poland, e-mail: edyta.pindelska@wum.edu.pl

**Abstract**

In this work we present crystal structure determination of two elusive polymorphs of furazidin, an antibacterial agent, employing a combination of crystal structure prediction (CSP) calculations and NMR crystallography approach. Two previously uncharacterized neat crystal forms, one of which has two symmetry independent molecules (form I), whereas the other one is a Z'=1 polymorph (form II), crystallize in $P2_1/c$ and $P-1$ space groups, respectively, and are both built by different conformers, displaying different intermolecular interactions. We demonstrate that the usage of either crystal structure prediction or NMR crystallography alone is insufficient to successfully elucidate the mentioned crystal structures, especially in the case of Z'=2 polymorph. In addition, cases of serendipitous agreement in terms of $^1$H or $^{13}$C NMR data obtained for the crystal structure prediction generated crystal structures different from the ones observed in laboratory (false-positive matches) are analyzed and described. Although for majority of analyzed crystal structures the obtained agreement with the NMR experiment is indicative of some structural features in common with the experimental structure, the mentioned serendipity observed in exceptional cases points to the necessity of caution when using NMR crystallography approach in crystal structure determination.

**Introduction**

Crystal structure prediction (CSP) of fairly rigid, small chemicals with one molecule in an asymmetric part of a unit cell has recently become almost a routine task (Price, 2018; Price *et al.*, 2016). This is mainly due to significant development of computational methods used in CSP in the last decades, leading to successful prediction of the most stable polymorphs of pharmaceutically relevant compounds (Bhardwaj *et al.*, 2019) and functional materials with desired properties (Pulido et al., 2017), explanation of crystallization behavior and preferred supramolecular synthons formed in crystals of organic molecules (Dudek *et al.*, 2019), including formation of solvates and hydrates (Braun *et al.*, 2016; Dudek & Day, 2019; Braun *et al.*, 2019), enhancing our understanding of gelation performance of molecules (Piana *et al.*, 2016), as well as crystal structure determination of active pharmaceutical ingredients, for which attempts to obtain appropriate crystal for single-crystal X-ray measurements have failed (Dudek *et al.*, 2018; Baias *et al.*, 2013b, Hofstetter *et al.*, 2019). Still, some challenges remain, including for example handling flexible molecules (Day *et al.*, 2017) and/or crystal structures with more than one molecule in an asymmetric part of a unit cell, as such CSP searches requires accounting for significantly more degrees of freedom, than the ones for Z'=1 and rigid molecules (Price, 2014; Reilly *et al.*, 2016).

The above mentioned challenges may be especially troublesome when using CSP as a method for crystal structure determination of polymorph(s) characterized before only by powder X-ray diffraction (PXRD), for which crystal structure is difficult to solve with standard analytical techniques (single-crystal X-ray measurements, solution from PXRD diffractogram). Then, if CSP search yield a structure which has noticeably lower energy than all other structures found in the same search, one may be fairly convinced that searched polymorph has been found. Much more often however, the number of structures found in the low-energy region of crystal energy landscape obtained from CSP search is significant, especially for Z'>1 searches, for which there can be even several thousands of crystal structures (Dudek *et al.*, 2017; Dudek *et al.*, 2016). This is still more problematic if we recall that an average error of the state-of-the-art methods used for lattice energy minimization is ca. 5 – 10 kJ/mol, a region in which one may often found tens of possible crystal structures. In such circumstances a comparison of all possible theoretical models with results from experimental methods (PXRD, solid-state NMR) is needed. On the other hand, these experimental methods can deliver structural constrains before any CSP search is attempted, thus considerably narrowing the number of degrees of freedom, which should be accounted for. In this work we demonstrate such an approach to crystal structure determination of two previously uncharacterized polymorphs of furazidin (Fig. 1), an active pharmaceutical ingredient (API) used for the treatment of urinary tract infections.

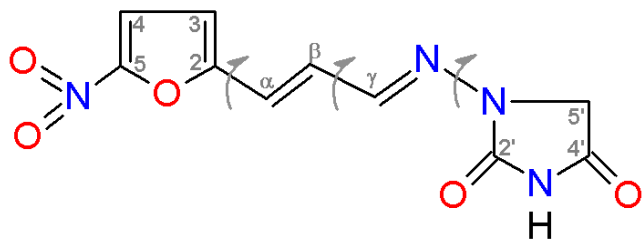

**Figure 1.** Chemical structure of furazidin with carbon atoms numbering. Arrows indicate rotatable bonds included in conformational search.

As will be shown, one of the studied polymorphs of furazidin crystallizes with two molecules in an asymmetric part of a unit cell (Z'=2), whereas the other one has only one symmetry unrelated molecule, a feature which enabled us a comparison of the level of difficulty in crystal structure determination using CSP-NMR approach between Z'=1 and Z'=2 polymorphs. This comparison includes the observed differences in extracting meaningful structural constrains from solid-state NMR and PXRD data, number of crystal structures found in the low energy region of CSP crystal energy landscape, as well as level of agreement with the experiment, especially in terms of NMR data. We also analyze cases of structures which display a very good agreement with the NMR experiment, despite being different from the actual experimental structure (false-positive indications).

**Materials & Methods**

*Crystallization experiments*

Powder sample of furazidin was obtained from Adamed Sp. z o. o., Czosnów, Poland. The phase purity of this polycrystalline sample was checked using PXRD and Fourier-transform infrared spectroscopy (FT-IR) as compared to the published data (patent no. WO 2015/181741 A1). These measurements have shown that it was form I of furazidin. Additionally form II was prepared using previously published procedures (patent no. WO 2015/181741 A1). The newly obtained form II has also been studied for phase purity. The obtained patterns and spectra were consistent with the published ones (FT-IR spectra are given in Figures S1 in Supporting Information, while both PXRD can be found in Figure 3). The high quality single crystals of DMF solvate monohydrate of furazidin were obtained by slow evaporation from dimethylformamide solution.

*Solid-state NMR measurements*

The majority of solid-state NMR measurements were performed with a Bruker Avance III 600 spectrometer, operating at 150.92 and 600.13 MHz for $^{13}$C and $^{1}$H, respectively. If not specified otherwise, samples were spun at 60 kHz using 1.3mm Bruker ZrO$_2$ rotors. Chemical shift were referenced with respect to finely powdered solid adamantane (1.8 ppm for $^{1}$H and 38.5 ppm for $^{13}$C). $^{1}$H-$^{13}$C *inv*HETCOR experiments were performed with a pulse sequence described by Pruski (Mao *et al.*, 2009; Althaus *et al.*, 2014). For both samples two experiments were measured with different second contact time equal to 50 μs and 2000 μs, first contact time was kept at duration 2000 μs. In the first case (short second contact time) only direct bond correlations are observed, while in the second case remote correlations are also visible. For $^{13}$C CP/MAS and $^{1}$H-$^{13}$C *inv*HETCOR experiments ramp shaped pulse from 90 to 100% was applied on the proton channel during CP with precisely optimized RF (on $^{13}$C, $^{15}$N labeled histidine hydrochloride) at around 160 kHz frequency. On the $^{13}$C channel an RF equal to 100 kHz during CP and π/2 pulses was applied. All experiments were performed with low power swept-frequency two-pulse phase modulation (SWf-TPPM) decoupling sequence (Chandran *et al.*, 2008) with an RF equal to 10 kHz. Repetition rate was set to 160 and 120 s for form I and II, respectively. $^{1}$H-$^{1}$H Back-to-Back correlation was measured with spinning speed equal to 62.5 kHz, with the duration of excitation and reconversion periods of 32 μs. To increase resolution in case of furazidin form I $^{1}$H-$^{13}$C *inv*HETCOR with short second contact time was measured at 800 MHz spectrometer with the acquisition parameters exactly the same as before. This was done to ensure appropriate separation of the correlation peaks, and as a result an unambiguous and precise assignment of the $^{1}$H resonances.

*Powder X-ray diffraction experiments*

Powder X-ray diffraction patterns were recorded at room temperature on a Bruker D8 Advance diffractometer equipped with a LYNXEYE position sensitive detector using Cu Kα radiation (λ = 0.15418 nm). The data were collected in the Bragg−Brentano (θ/θ) horizontal geometry (flat reflection mode) between 3° and 45° (2θ) in a continuous scan using 0.03° steps and 384 s/step. The diffractometer incident beam path was equipped with a 2.5° Soller slit and a 1.14° fixed divergence slit, while the diffracted beam path was equipped with a programmable anti-scatter slit (fixed at 2.20°), a Ni β-filter, and a 2.5° Soller slit. Data were collected under standard laboratory conditions.

Indexing of powder X-ray diffractograms was performed with Reflex tool, as implemented in Material Studio software, using X-Cell (Neumann, 2003) and searching through all possible crystal systems. Rietveld refinement was performed for the selected structures from CSP search with GSAS-II software (Toby & Dreele, 2013), allowing cell parameters to be refined, while keeping atomic positions fixed.

*FT-IR Spectroscopy measurements*

FT-IR studies were performed using a Perkin-Elmer Spectrum 1000 FT-IR spectrometer. Samples were mixed with KBr using a mortar and a pestle to obtain a homogenous mixture, then the powder was gently pressed under vacuum conditions with a compensation force of 10 ton using 14 mm diameter round flat force punch to produce a KBr pellet. Samples were placed in the light path and the IR spectra from 400 to 4000 cm$^{-1}$, resolution 2 cm$^{-1}$, in a transmission mode were registered. All spectra were prepared using GRAMS AI (version 8.0, Thermo Electron Corp., Waltham, MA, USA).

*DSC measurements*

DSC measurements were performed on a differential scanning calorimeter DSC 2920, TA Instruments, in the temperature range of 273 – 573 K (0 – 300 °C), with a heating rate of 2 K/min, nitrogen flow of 50 mL/min, and using aluminum pans. Temperature calibration was done in a two-point mode on indium and tin, and the data were acquired and analyzed with Thermal Analyst software. DSC results available in Supporting Information (Figure S2).

*Conformational search & Crystal Structure Prediction calculations*

Conformational search has been performed using Conformers tool as implemented in Materials Studio software, using all three rotatable bonds with 60° step for each bond. Subsequently all conformers has been optimized using universal force field, which yielded 22 different structures. Out of these 8 unique conformers were obtained after final optimization using Gaussian16 software (Frisch *et al.*, 2016) and B3LYP functional (Becke, 1993) supplemented by the Grimme D3 dispersion correction using Becke-Johnson damping (GD3BJ) (Grimme *et al.*, 2011) and 6-311G(d,p) basis set.

For crystal structure prediction calculations of Z' = 1 polymorph (form II) four conformers selected on the basis of NMR experiment were used. For each conformer 10000 successfully lattice energy minimized crystal structures were generated in each of the selected space groups, *i.e.* P2$_1$/c, C2/c, P-1 and P1. The space groups selection was made on the basis of indexing of experimental powder X-ray diffractogram. Crystal structures were generated using Global Lattice Energy Explorer code (Case *et al.*, 2016) and lattice energy minimized with DMACRYS 2.2.1.0 software (Price *et al.*, 2010), based on atom-centered distributed multipoles up to hexadecapoles, calculated for each conformer independently from B3LYP-D3BJ/6-311G** charge density using distributed multipole analysis (DMA) in GDMA 2.2.11 software (Stone, 2005). The electrostatic interactions were calculated as a sum of charge – charge, dipole – dipole and higher multipoles interactions. Intermolecular interactions (repulsion-dispersion term) were calculated using FIT potential (Coombes *et al.*, 1996), with 25 Å cut-off on van der Waals interactions. To eliminate duplicates the obtained crystal structures were clustered on the basis of their energy, density (with a tolerance of 0.1 kJ/mol and 0.02 g/cm$^3$, respectively) and similarity of PXRD pattern. Final total energy for each structure is given as a sum of intra- and intermolecular energy contributions. The first term is the relative gas phase energy of conformers in respect to the gas phase global minimum, calculated at B3LYP-D3BJ/6-311G** level of theory, whereas the second one is the lattice energy term obtained from DMACRYS lattice energy minimization.

Crystal structure prediction of Z' = 2 polymorph of furazidin (form I) required a more extensive search, than the one performed for form II. Therefore it was done in two stages. First 10000 crystal structures were generated for each of 36 possible pairs of conformers in P2$_1$, Pc and P-1 space groups, using the same methodology as given above. Then, for the conformes which yielded the lowest energy structures the search has been repeated, generating this time 50000 structures. Also, an additional CSP search for the pair of conformers giving the best agreement in terms of NMR experiment was performed in the 11 most common space groups (apart from the ones for which a search was already done before), *i.e.* P2$_1$/c, C2/c, P2$_1$2$_1$2$_1$, Pbca, Pna2$_1$, Cc, Pnma, P1 and C$_2$. The plots presented in the manuscript shows CSP search results from the first stage (with generation of 10000 crystal structures), while all the other results are given in Supporting Information (Figures S5 and S6).

*DFT-D2 energy re-ranking of crystal structures and calculations of NMR parameters*

In each case CSP search was followed by DFT-level optimization using PBE functional and Grimme-D2 dispersion correction scheme (Grimme, 2006), as implemented in CASTEP (Clark *et al.*, 2005). First, all atomic positions were allowed to relax, while unit cell parameters were kept rigid, then also unit cell parameters were allowed to relax. For each optimization 600 eV energy cut-off on plane wave basis set and the k-point spacing of 0.07 Å$^{-1}$ were used. These parameters had been primarily tested for convergence. The number of structures selected for DFT optimization was equal to 14 and 42 for form II and form I, respectively. For the selection criteria see Results & Discussion part.

After DFT optimization NMR parameters were calculated using GIPAW methodology (Pickard & Mauri, 2001; Yates *et al.*, 2007) and the same level of theory as described above. The obtained theoretical shielding constants ($\sigma_{calc}$) were recalculated to chemical shifts ($\delta_{calc}$) using the following equation: $\delta_{calc} = (\sigma_{calc} - b) / m$, where b and m are the intercept and slope obtained after plotting experimental chemical shifts and theoretical shielding constants and calculating linear regression curve. Subsequently, root-mean-square (RMS) deviation was calculated for each data set.

*Structural analysis and visualization software*

The molecular Hirshfeld surfaces (HS) (Spackman & Jayatilaka, 2009; McKinnon *et al.*, 2007) were calculated from crystal structure coordinates using CrystalExplorer software (Turner *et al.*, 2017) and used to highlight the differences in the molecular environments. The surfaces were constructed based on the electron distribution calculated as the sum of spherical atom electron densities. The normalized contact distance ($d_{norm}$) based on both $d_e$ (the distances from the surface to the closest atom outside the surface) and $d_i$, (the distances from the surface to the closest atom inside the surface) and the vdW radii ($r$) of atoms is given by the equation: $d_{norm} = [(d_i - r_i)/ r_i]+[(d_e - r_e)/ r_e]$. To visualize molecular packing in the analyzed crystal structures both CrystalExplorer 17 (Turner *et al.*, 2017) and Chimera 1.12 (Pettersen *et al.*, 2004) software were used.

The interaction energy calculations were performed using the CrystalExplorer 17 software (Turner *et al.*, 2017) using crystal structures from the obtained cif files as inputs. In each case a cluster with radius of 3.8 Å around a molecule present in the asymmetric unit was generated. The neighboring molecules in the shell around the central molecule were generated by applying crystallographic symmetry operations. The interaction energy calculations ($E_{tot}$) between symmetry independent molecules in the crystalline environment were obtained from monomer wavefunctions calculated at the CE-B3LYP/6-31G(d,p) level of theory with the following scaling factors to determine $E_{tot}$: $k_{ele}= 1.057$, $k_{pol} = 0.740$, $k_{dis} = 0.871$, and $k_{rep} = 0.618$ (Mackenzie *et al.*, 2017). The energies are expressed in terms of electrostatic ($E'_{ele}$), polarization ($E'_{pol}$), dispersion ($E'_{dis}$) and exchange repulsion ($E'_{rep}$) terms. The total energy ($E_{tot}$) is represented as $E_{tot} = k_{ele} E'_{ele} + k_{pol} E'_{pol} + k_{dis} K'_{dis} + k_{rep} K'_{rep}$, where k values are the scale factors. The magnitudes of the intermolecular interaction energies were graphically represented using energy frameworks, in which the topology of interactions may be conveniently described as a network of tubes connecting centers of molecules. The tube width is proportional to the energy of interaction, while its color depends on the nature of interaction energy. In Tables S3-S5 energy frameworks corresponding to the different energy components and the total interaction energy in both furazidin forms are presented.

**Results & Discussion**

*Preliminary experimental results for form I and II of furazidin*

So far the only known crystal structures of furazidin are its two solvates – monohydrates, one with tetrahydrofuran (CCDC refcode ASATIZ) and one with N,N-dimethylformamid (CCDC refcode ASATOF) (Berzins *et al.*, 2016), but patent literature indicates that there are two neat polymorphs of this API (patent no. WO 2015/181741 A1). Our crystallization attempts did not yield any crystallographic-quality monocrystals, resulting only in microcrystalline powders. Their identities with forms I and II of furazidin were confirmed with powder X-Ray diffraction.

The $^{13}$C CPMAS spectra of both polymorphs, as well as that of DMF solvate – monohydrate (Figure 2) suggest, that form I has two molecules in an asymmetric part of a unit cell, which is visible as doubling of almost all observed $^{13}$C resonances. Form II, on the other hand, seem to be very similar to DMF solvate – monohydrate in terms of $^{13}$C NMR chemical shifts, with only one symmetry unrelated molecule. Contrarily, both neat forms display very similar thermal behavior, with decomposition observed at temperatures 289 and 288°C for form I and II, respectively (for DSC plots see Supporting Information, Figure S2).

*Conformational search*

Furazidin is fairly rigid molecule with only three rotatable bonds marked with grey arrows in Figure 1. Its conformational search yielded 22 conformers, which after re-optimization at the DFT level of theory gave final 8 distinct conformers with the highest relative gas phase energy in respect to the best one being equal to 22.5 kJ/mol. This means that all 8 conformers can form stable crystal structures (Thompson & Day, 2014).

*Extracting experimental constrains*

Powder X-ray diffractograms of forms I and II are given in Figure 3. Indexing of PXRD diffractogram of form II yielded three plausible unit cells, one belonging to monoclinic, and two other to triclinic crystal systems. The first unit cell with parameters being equal to a = 11.62 Å, b = 11.42 Å, c = 8.48 Å and β = 106.91° should contain four molecules of furazidin to yield probable density of 1.63 g/cm$^3$, which means four symmetry operations. Hence the most probable space groups are Cm, Cc, P2/c, C2/m, P2/m, P2$_1$/m, P2$_1$/c, C2/c. Of these P2$_1$/c and C2/c are the most common ones, and their Pawley refinement against the experimental diffractogram yielded Rwp equal to 8.47%. The unit cells belonging to triclinic crystal system gave somewhat worse agreement with PXRD experiment, with Rwp values obtained after Pawley refinement equal to 9.63% and 10.0% for cells with parameters a = 11.49 Å, b = 8.59 Å, c = 8.52 Å, α = 66.95°, β = 104.79°, γ = 96.62° (d = 1.17 g/cm$^3$) and a = 11.13 Å, b = 8.46 Å, c = 8.13 Å, α = 104.37°, β = 90.43°, γ = 92.96° (d = 1.18 g/cm$^3$), respectively. As a result it seems the most reasonable to do a CSP search first in P2$_1$/c and C2/c space groups, followed by a search in P1 and P-1 space groups.

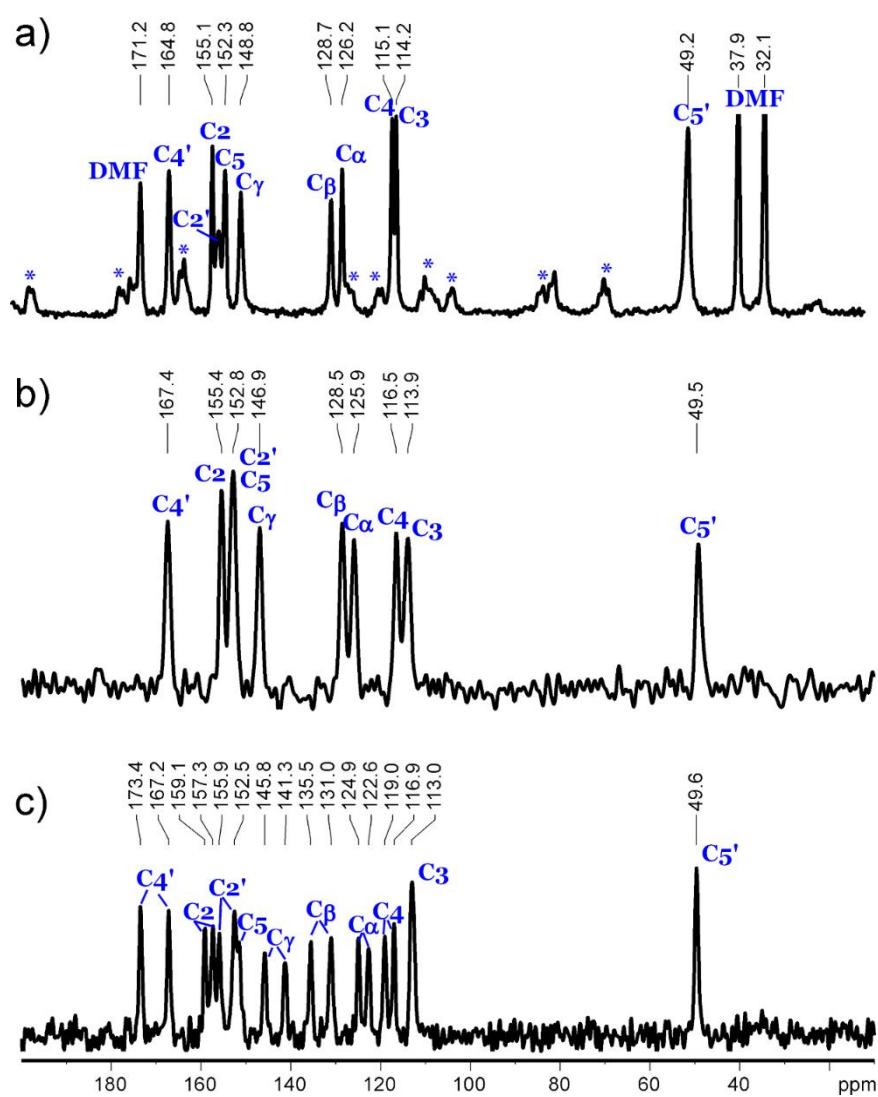

**Figure 2.** $^{13}$C CPMAS NMR spectra with $^{13}$C signal assignment of three crystalline forms of furazidin: DMF solvate monohydrate (a), form II (b), and form I (c). Asterisks in the DMF solvate spectrum mark spinning sidebands. Note, that this spectrum was registered with 10

kHz spinning speed, whereas spectra shown in b) and c) with 60 kHz spinning speed, and so spinning sidebands for these spectra are outside the observed spectral region.

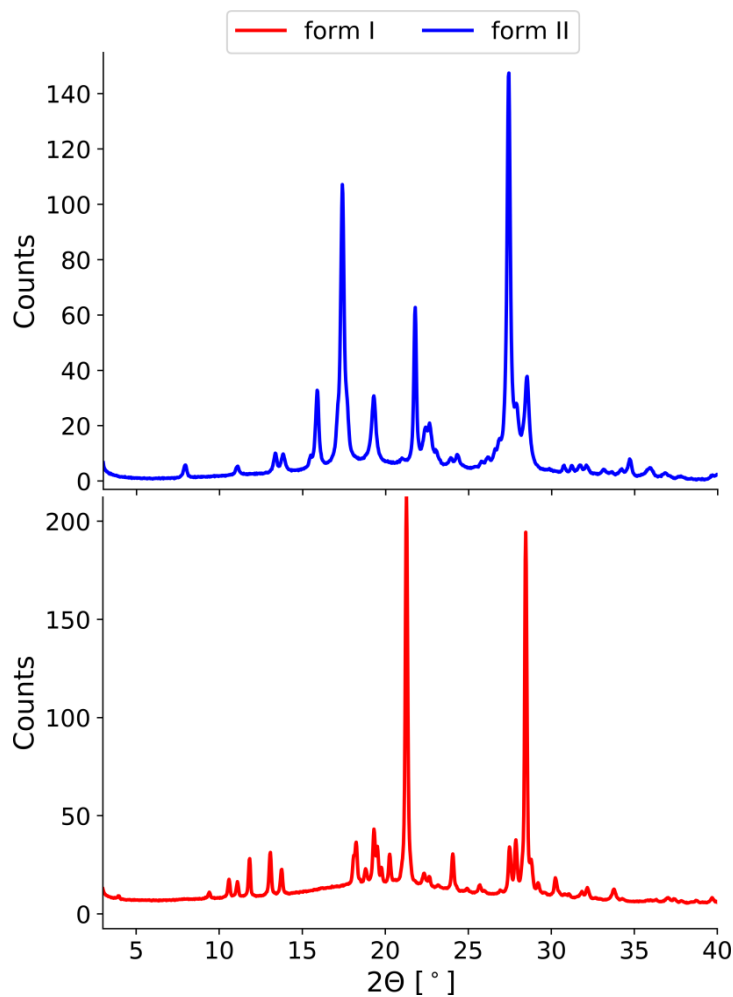

**Figure 3.** Powder X-ray diffractograms for form I and II of furazidin.

In the case of form I PXRD diffractogram indexing gave two possible unit cells, in monoclinic and triclinic crystal systems. The obtained cell parameters for the mentioned crystal systems were equal to a = 22.63 Å, b = 6.54 Å, c = 9.72 Å, β = 99.47°, which corresponds to the density of 1.24 g/cm$^3$ for 4 furazidin molecules, and a = 10.79 Å, b = 9.98 Å, c = 8.33 Å, α = 64.98°, β = 110.13°, γ = 117.27°, corresponding to the destiny of 1.24 g/cm$^3$. The Rwp values for both unit cells obtained after Pawley refinement were equal to 5.54 and 10.8%, respectively. Hence the most probable space groups for this polymorphic form are P2, P2$_1$, Pm, Pc, P1 and P-1.

The next experimental constrain, which can significantly facilitate CSP search is the conformation of a compound in a crystal. One of the best ways to extract constrains enabling determination of a conformer is *via* solid-state NMR measurements under very fast magic angle spinning (VF MAS), *i.e.* with the spinning speed of at least 42 kHz. In such conditions it is usually possible to reliably assign all $^1$H and $^{13}$C resonances, as well as to determine additional $^1$H-$^{13}$C or $^1$H-$^1$H through-space correlations, which arise from proximities of certain atoms in a molecule or in a crystallographic unit cell. Figure 4 shows *inv*-$^1$H-$^{13}$C HETCOR correlation spectra with long contact time, registered at 60 kHz spinning speed for both polymorphs, while their $^1$H and $^{13}$C chemical shifts can be found in Table S1 in Supporting Information. $^1$H and $^{13}$C signals assignment for form II is quite straightforward task and does not create any difficulties. In addition a correlation between Cγ and H5'a indicates that these atoms must be close in space, a condition which is fulfilled only by 4 out of 8 conformers.

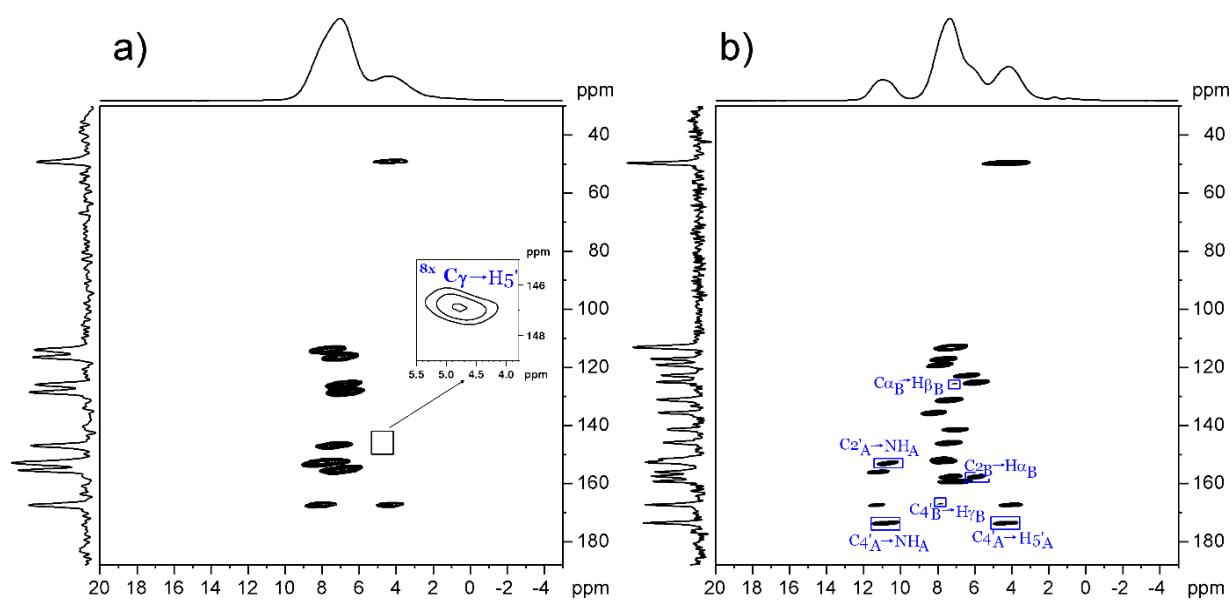

**Figure 4.** *inv*-$^1$H-$^{13}$C HETOCR spectra with long contact time registered at the spinning speed of 60 kHz for form II (a) and I (b) with assigned crucial correlation peaks.

In the case of form I, the analysis of its 2D spectra proves that in form I there are indeed two symmetry independent molecules, as suggested before by one-dimensional NMR measurements. The observed $^1$H-$^{13}$C correlations allow to assign all resonances to respective atoms in furazidin molecules, but there are some ambiguities as to the full assignment to each molecule separately. For example, it is possible to allocate resonances originating from C2', C4', C5', Cγ, Hγ, H5' and NH to molecules A and B, as well as to identify correlations between Cα, Cβ, C2, C3, C4, C5 and their corresponding protons, but the correlations which could help to combine those two groups together and thus enable the full assignment to molecules A and B are inconclusive due to signals overlap. As a result we decided to check parameters calculated for theoretical structural models against two possible combinations of the assignment. As for additional structural constrains there is a weak correlation, which can be assign to C5' and Hγ, suggesting that again, the same 4 conformers should be accounted for primarily. However, in this case it is also possible that this correlation originates from different proton than Hγ, as its resonating frequency is close to those found for other $^1$H atoms. It is also unclear whether this structural constrain should be used for one or for both molecules. As a result it seems more reasonable not to exclude any conformer and do CSP search for all possible combinations of conformers (*i.e.* 36 combinations), but limit the searched space groups from among those indicated by PXRD indexing to the most common ones, that is P-1, P2$_1$ and Pc.

Apart from the structural constrains, which can be used in the selection of plausible conformers, solid-state NMR spectra can deliver information on the intermolecular close contacts. In majority of cases it is impossible to distinguish a priori which correlations arise from intramolecular interactions (information on the conformation of a molecule), and which from intermolecular ones (information of mutual arrangement of molecules inside a crystal). However, this is not the case if there are two molecules in an asymmetric part of a crystallographic unit cell and visible correlation signals arise from the same nuclei in distinct molecules. Obviously, such correlations can only be the result of intermolecular interactions, thus giving additional structural constrains, which cannot be used before a CSP search, but can allow for the elimination of the CSP-generated model structures, which do not fulfill these constrains. This is the case of furazidin form I. In its $^1$H-$^1$H Back-to-Back correlation spectrum there is a clearly visible correlation between two protons resonating at 10.8 and 11.3 ppm (see Figure S4, Supporting Information). These signals originate from amide (NH) protons in molecules A and B, respectively. As a result, these two protons have to be close in space in the crystal of form I.

Summarizing this part, we were able to successfully narrow down the number of possible space groups, which should be accounted for in a CSP search for form I and II of furazidin, as well as to narrow down to four the number of possible conformers that can be present in the crystal structure of form II. In addition we were able to extract an additional structural constrain for form I, which can be used after the CSP search, *i.e.* close proximity of NH$_A$ and NH$_B$ protons.

*CSP search for form II (Z'=1)*

The CSP crystal energy landscape for four selected conformers in P2$_1$/c, C2/c, P1 and P-1 space groups is shown in Figure 5. The lowest energy region of this plot, that is the first 20 kJ/mol above the global minimum, is very crowded with more than 500 distinct crystal structures. Eight lowest energy structures including the landscape's global minimum, are built by a conformer with the lowest gas-phase energy among the tested ones, with intermolecular energy 8.94 kJ/mol above the global gas phase minimum. All the other conformers build significantly less favorable structures, with the best one being 5 kJ/mol higher in energy than the global minimum structure. Noteworthy, all structures built by the highest-energy conformer (with gas phase energy 22.5 kJ/mol higher than the gas phase global minimum, marked with red in Figure 5) are very high in total energy, with the best one having its total energy 18 kJ/mol higher than the CSP landscape global minimum. This suggest that all conformers of furazidin are able to pack quite efficiently in a crystal structure, and the factor that determines their final energy is primarily intramolecular energy contribution (*i.e.* gas phase energy of a conformer).

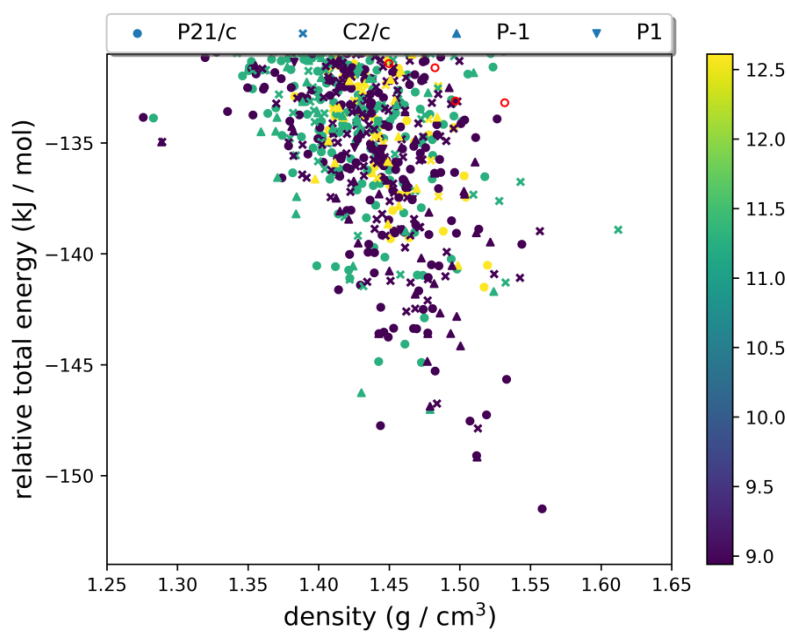

**Figure 5.** CSP energy landscape for four possible conformers tested in $P2_1/c$, $C2/c$, $P1$ and $P\text{-}1$ space groups for Z' = 1 polymorph of furazidin (form II). Structures obtained for different conformers are color-coded according to their gas phase energy in kJ/mol, as shown in a colorbar, except for the highest-energy gas phase conformer, which is shown in red.

As for the space group the most energetically favorable structures seem to be those in $P2_1/c$ space group, but those found in C2/c and P-1 are only slightly less favorable. The global minimum structure, with total energy only 1.5 kJ/mol lower than the energy of the second best structure, has unit cell parameters equal to a = 11.66 Å, b = 11.94 Å, c = 8.47 Å and β =

107.22°, in close agreement with one of the possible unit cells indicated by indexing powder X-ray diffractogram. However, to further evaluate whether this structure is the one observed experimentally, it needs to be re-optimized, having NMR parameters calculated and compared with experimental ones. In addition, we decided to calculate NMR parameters also for the lowest-energy structures of each conformer in each of the tested space groups to check the level of agreement for as wide variety of structures as possible and determine whether there will be a significant difference in the agreement between this structure and all the other.

The re-optimization of the selected model structures at the DFT-D2 level of theory resulted in changes in the final energy ranking. After re-optimization the force field global minimum structure was now found to be third best structure, with relative total energy 1.7 kJ/mol higher than DFT-D2 global minimum. This energy difference, however is well within error boundaries of the applied dispersion correction scheme. As for the $^1$H and $^{13}$C RMS values, which express the level of agreement between experimental and theoretical NMR parameters, the force field global minimum structure is unquestionably the one showing the best agreement with the NMR experiment, with 0.3 and 2.6 ppm for $^1$H and $^{13}$C RMS, respectively (see Figure 6). This confirms our previous assumption that this structure is the one observed experimentally. Final proof is delivered by Rietveld refinement of this structure against the experimental powder X-ray diffractogram, which yielded a very good Rwp value of 9.60 % (for the comparison of simulated and experimental PXRD diffractograms see Supporting Information, Figures S7 and S8). Note, that from now on, throughout the manuscript we refer to this structure as to the experimental one for form II. By doing so we mean that this is the structure actually observed in laboratory for form II of furazidin, but it was obtained *via* theoretical methods and then confirmed and refined by the performed solid-state NMR and PXRD experiments.

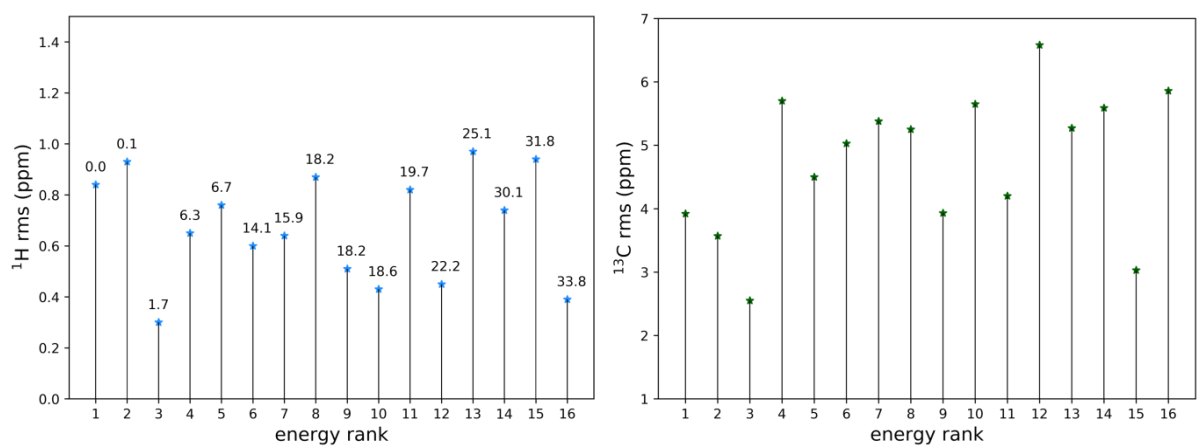

**Figure 6.** $^1$H and $^{13}$C rms values (in ppm) obtained after comparison of experimental and theoretical NMR parameters for the selected model crystal structures obtained from CSP search for 4 conformers in P2$_1$/c, C2/c, P-1 and P1 space groups. The annotated values in $^1$H rms plot represent relative total DFT-D2 energies of the model crystal structures (in kJ/mol, calculated per one furazidin molecule) in respect to the lowest energy structure.

It is worth to note that the structure indicated as the one of form II is actually the only one having low RMS values for both tested nuclei, while some of the other model crystal structures have quite low RMS value for $^1$H only or $^{13}$C only. For example, the 16$^{th}$ structure (according to its DFT-D2 energy rank), with relative total energy as high as 33.8 kJ/mol above the DFT-D2 global minimum, has surprisingly low $^1$H RMS value, equal to 0.39 ppm. This structure, however, apart from being too far from the energetic global minimum to be designated as the experimental one, shows very poor agreement with the experiment in terms of $^{13}$C NMR values, with an RMS equal to 5.8 ppm. In our previous studies we have shown that the sensitivity of $^1$H NMR parameters to local changes in chemical environment in a crystal is usually significantly higher that the sensitivity of $^{13}$C NMR data, and is indicative of a molecular conformation inside the crystal, rather than crystal packing itself (Dudek *et al.*, 2017; Dudek *et al.*, 2016). In the case of furazidin, however, looking only at $^1$H NMR data seems to be not enough to unambiguously indicate the correct structural model, and this may be due to at least two factors. First is the presence of two carbonyl groups, which chemical shifts are very sensitive to intermolecular interactions, and the second one is the fact that furazidin has limited conformational freedom, with the obtained conformers being similar to each other in terms of chemical environment around particular nuclei. As a result it is more probable that serendipitous agreement with the experiment will be observed for one of the regarded nuclei, but it is unlikely that this will be the case for both nuclei ($^1$H and $^{13}$C).

Having this in mind we decided to examine all cases of false-positive indications in terms of $^1$H RMS values, that is all model crystal structures with low $^1$H RMS value, but not being the one observed experimentally. Apart from the experimental crystal structure, there are four model crystal structures with $^1$H RMS value equal to or lower than 0.5 ppm, a level of agreement which in case of a lack of all the other data might be considered as enough to prove an identity with the experimental crystal structure (Hofstetter & Emsley, 2017; Baias *et al.*, 2013a). These are 9$^{th}$, 10$^{th}$, 12$^{th}$ and 16$^{th}$ structures with $^1$H RMS values equal to 0.51, 0.43, 0.45 and 0.39 ppm, respectively. Only the 9$^{th}$ structure, found in P1 space group, is built by the same conformer as the experimental structure and it has 7 out of 15 molecules in common with the experimental structure with an RMS value of 0.148 Å, when compared with Crystal Packing Similarity Tool (Chisholm & Motherwell, 2005), using default comparison parameters (15-molecule cluster, geometrical tolerance of 20% for atomic distances and 20° for angles). In both structures furazidin molecules form layers held together by N-H...O intermolecular hydrogen bonds, and in both cases furazidin molecules are arranged in the same way within each layer, but there is a different arrangement of the layers in respect to each other. The other three structures with low $^1$H RMS are built by different conformers than the experimental structure, showing that in case of furazidin the agreement in terms of $^1$H NMR data is not necessarily indicative of a conformation inside the crystal. In contrast, in all of them furazidin molecules interact *via* N-H...O intermolecular hydrogen bonds, a feature not observed in crystal structures with much worse agreement in terms of $^1$H NMR data. On the other hand, the same hydrogen bonding interactions seem to exhaust the list of similarities between the compared crystal structures. As an example, Figure 7 shows the arrangement of furazidin molecules in the experimental crystal structure and the 16$^{th}$ crystal structure, which is the one with the next best agreement in terms of $^1$H NMR data. In this case both furazidin molecules within one layer, as well as mutual arrangement of layers interacting *via* short contacts is different for both structures, which finds a reflection in a disagreement in terms of $^{13}$C NMR data for the 16$^{th}$ structure.

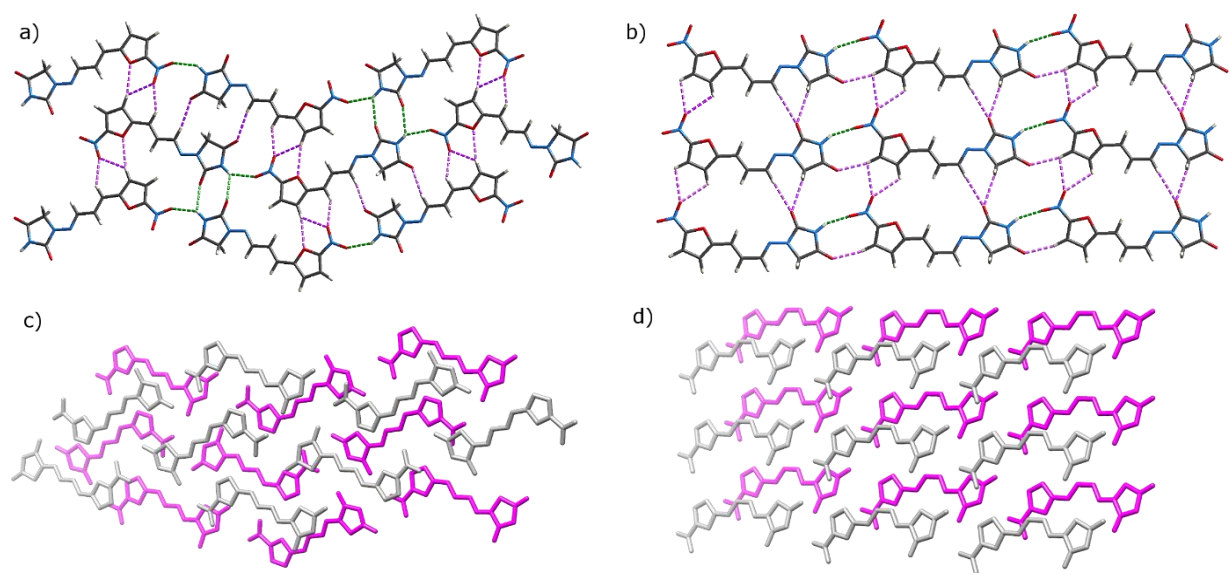

**Figure 7.** Molecular arrangement of furazidin molecules in Z' = 1 polymorph (form II): the experimental crystal structure (a and c) and 16$^{th}$ (according to its DFT-D2 energy rank) theoretical crystal structure (b and d). The upper panel (a and b) shows the arrangement within one layer, N-H...O hydrogen bonds are marked with green dotted lines, short C-H...O

contacts with purple dotted lines; the lower panel (c and d) shows the inter-layer arrangement, with two interacting layers represented by different colors.

As a summary, it can be concluded that the agreement in terms of $^1$H solid-state NMR data alone may in some cases be serendipitous (though it usually is associated with some similarities with the experimental structure), and as such is not enough to prove the identity of the examined crystal structure. However, if this indeed is the case, it should reflect itself in a disagreement in terms of $^{13}$C NMR data. Lastly, as it is difficult to decide *a priori* how good agreement in terms of the NMR data is good enough, it is necessary to account also for relative total energy of a regarded structure. Its value indicates the most favorable (and therefore the most probable) arrangement of molecules inside a crystal lattice. Therefore, a model crystal structure with good agreement in terms of NMR parameters cannot be regarded as the experimental one if its relative total energy is significantly higher that some of the other structures.

### *CSP search for form I (Z' = 2)*

The computational search for the crystal structure of form I is a much more demanding task than the one for form II. As pointed out in the *Extracting experimental constrains* section it was not possible to exclude any conformation on the basis of experimental constrains (with only slight indication to a conformation similar to the one found in crystal structure of form II), which means that 36 possible combinations of all 8 conformers had to be tested at least in three selected space groups, *i.e.* in P2$_1$, Pc and P-1. The lowest energy region of the resulting CSP crystal energy landscape is shown in Figure 8. In this case, similarly as it was observed for Z' = 1 polymorph, the majority of low energy structures are built by low energy gas phase conformers, although here the effect is not as strong as it was the case of form II. Clearly, the formation of Z' = 2 crystal structure can result in stronger intermolecular interactions, thus enabling building Z' = 2 structures by less energetically favored conformers. As to the space group preferences, it seems that there is almost an equal distribution of structures in all three tested space groups in the lowest energy region of the landscape.

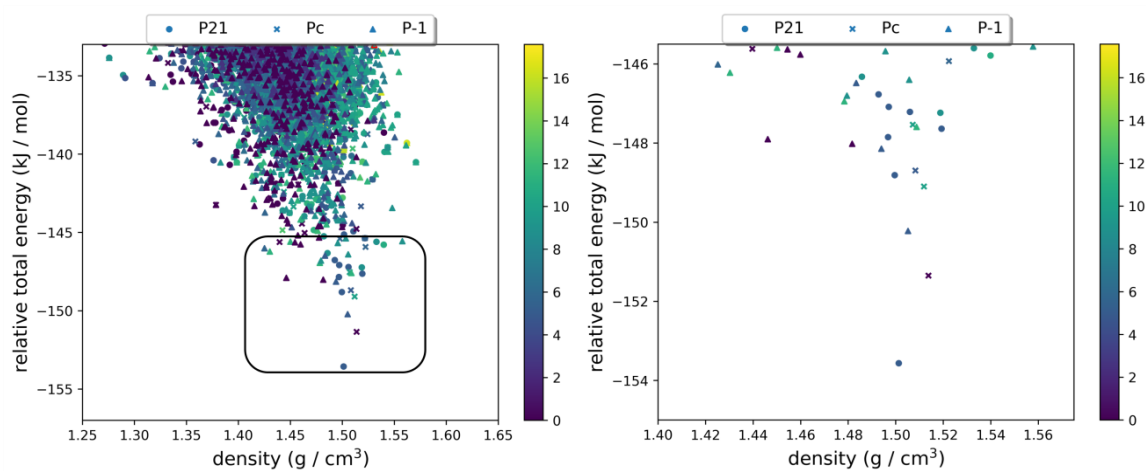

**Figure 8.** CSP energy landscape for all possible combinations of conformers tested in P2$_1$, Pc and P-1 space groups for Z' = 2 polymorph of furazidin (form I). Structures obtained for different conformers are color-coded according to their gas phase energy in kJ/mol (which is an averaged value for two conformers present in each structure), as shown in a colorbar. Black rectangle indicates the expansion region shown at the right-hand side of the figure.

The observed global minimum on the force field landscape belongs to the structure found in P2$_1$ space group, with unit cell parameters equal to a = 9.62 Å, b = 11.26 Å, c = 11.21 Å and β = 102.79°, different from those indicated by indexing of experimental PXRD diffractogram for form I of furazidin. Therefore, this time global minimum structure is probably not the one observed experimentally. In a search for the experimental crystal structure a comparison with NMR experiment of the most plausible candidates would be of use. However, the low-energy region of the CSP landscape for Z' = 2 polymorph is even more crowded than it was observed for form II, with over 3600 structures having relative total energy of less than 20 kJ/mol above the global minimum. Obviously, re-optimization of such a number of structures with DFT-D2 method would consume prohibitively large amount of computational time. Instead, we decided to re-calculate at this level of theory only the lowest energy structures for each conformer in each of the tested space groups which had total relative energy lower than 15 kJ/mol above the global minimum. This procedure, followed by NMR parameters calculations, should indicate which of the tested structures are close in terms of NMR data to the experimentally observed structure. Then, if needed, more exhaustive search for the most promising candidates can be performed.

The above presented criteria resulted in the selection of 42 possible crystal structures, built by 18 different combination of conformers. As expected, their re-optimization somewhat changed the energy ranking, so that now the DFT-D2 global minimum structure is the one found in P-1 space group, with unit cell parameters of a = 22.44 Å, b = 12.45 Å, c = 5.33 Å, α = 50.96°, β = 92.50°, γ = 97.02°. These parameters bear some resemblance to the ones obtained for monoclinic crystal system from indexing of PXRD diffractogram for form I (see section *Extracting experimental constrains*). Rietveld refinement performed for this structure yielded a very good agreement of Rwp = 13.23% (for the comparison of experimental *vs.* simulated PXRD diffractogram see Supporting Information, Figure S8). As a result this structure can be considered the experimental one (the one observed in laboratory for form I of furazidin), provided that it will show a very good agreement with the NMR experiment.

Figure 9 presents the $^1$H and $^{13}$C RMS values obtained after comparison of theoretical and experimental NMR data for 42 model crystal structures. Indeed, the global minimum structure shows the best agreement in terms of $^1$H chemical shifts, with an RMS value equal to 0.41 ppm and the second best in terms of $^{13}$C chemical shifts (RMS = 2.45 ppm). Note, that in terms of $^1$H RMS value this agreement seem to be worse than the one obtained for form II of furazidin, but this is not the case when looking at the correlation coefficient ($R^2$) obtained from linear regression. In fact, the $R^2$ value for furazidin form I is equal to 0.963, while for form II it is equal to 0.955 (for $R^2$, RMS, slope and intercept values of all the calculated structures see Table S4, Supporting Information). This structure is built by a pair of identical conformers, forming dimers interacting *via* hydrogen bonds (Figure 10) through C-H...O bonds. As a result the NH protons of both symmetry unrelated molecules are in close proximity, with a distance of only 2.52 Å. Therefore, this structure fulfills also the intermolecular structural constrain extracted from $^1$H-$^1$H BaBa NMR spectrum. Noteworthy, the conformer found in this structure belongs to the same group of conformers as the one found in form II of furazidin, which was also suggested by the results of solid-state NMR measurements. Given all the above evidence, we conclude that the DFT-D2 global minimum structure is the experimental structure of form I. From now on we will refer to this structure as to the experimental one, as was justified previously for the structure obtained for form II.

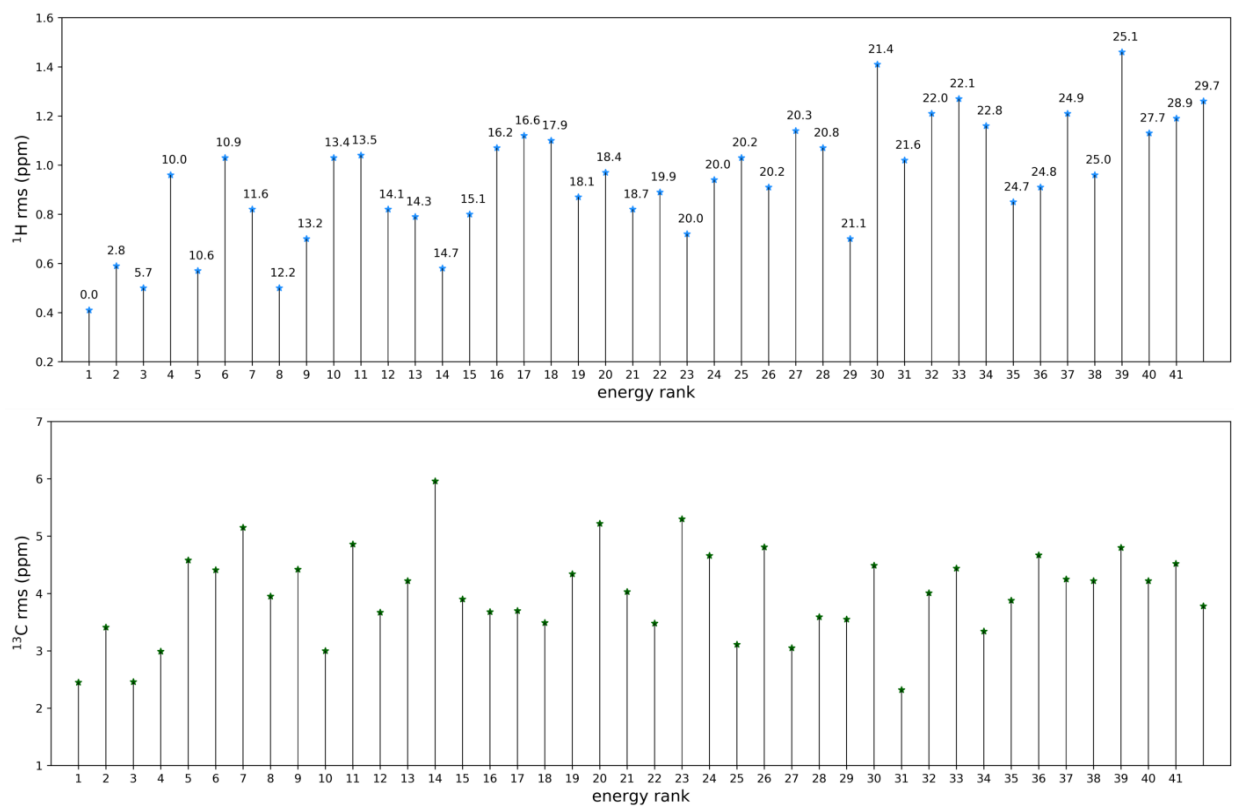

**Figure 9.** $^1H$ and $^{13}C$ rms values (in ppm) obtained after comparison of experimental and theoretical NMR parameters for the selected model crystal structures obtained from CSP search for furazidin Z' = 2 polymorph (form I) – the search has been performed for 36 conformers in $P2_1$, Pc and P-1 space groups. The annotated values in $^1H$ rms plot represent relative total DFT-D2 energies of the model crystal structures (in kJ/mol, calculated per one furazidin molecule) in respect to the lowest energy structure.

As a side note, it is worth to point out that indexing of PXRD diffractogram of form I did not yielded the unit cell, which was eventually found for the experimental structure, but only a similar one, found in a different crystal system. This is not surprising if we account for the fact that molecules, which do not want to produce crystallographic-quality monocrystals often have a tendency to show rather poor degree of crystallinity, a feature which can significantly influence the quality of PXRD data.

As before, it is worthwhile to look at all the other model crystal structures giving good agreement in terms of NMR parameters. Among 42 tested structures there are six structures having $^1H$ RMS value lower than 0.6 ppm: the experimental crystal structure, 2[nd], 3[rd], 5[th], 8[th] and 14[th] crystal structures (numbering of structures is made according to their DFT-D2 energy ranks). Of these only two crystal structures have also good agreement in terms of $^{13}C$ NMR data, with the RMS value lower than 3.0 ppm: the experimental structure and the 3$^{rd}$ crystal structure. This latter structure, found in P2$_1$ space group, is built by exactly the same pair of conformers as the experimental one, displays similar, though not identical, hydrogen bonding pattern (Figure 10b) and has 6 out of 15 molecules in common with the experimental structure (RMS = 0.456 Å), when compared using Crystal Packing Similarity Tool and default comparison parameters. It also fulfills the condition of close proximity of NH$_A$ and NH$_B$ protons (with the N-H...N intermolecular distance equal to 2.99 Å). These similarities explain the agreement in terms of NMR parameters observed for this structure. Out of the remaining four model crystal structures having good agreement in terms of $^1$H NMR data, three contain at least one conformer identical with the experimental crystal structure, and the 8$^{th}$ crystal structure display the same hydrogen bonding pattern, fulfills the NH$_A$ and NH$_B$ proximity condition, and has 6 out of 15 molecules in common with the experimental structure (RMS = 0.626 Å). Only the 14$^{th}$ crystal structure is different from the experimental structure in terms of both conformers building this structure and its hydrogen bonding pattern. This manifests itself in a poor agreement in terms of $^{13}$C NMR data, which actually is the worst from among the tested structures and equal to 5.96 ppm.

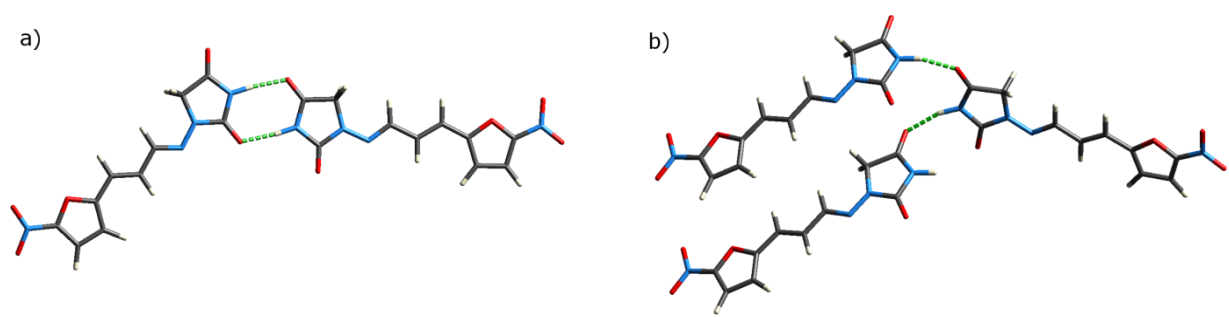

**Figure 10.** Hydrogen bonding pattern in the experimental crystal structure of form I (a) and in the third lowest energy crystal structure (b), built by the same conformers as the experimental structure; hydrogen bonds are marked with dotted green lines.

To conclude this part, in the case of form I the agreement in terms of $^1$H NMR data is more indicative of similarities of the regarded model crystal structures with the experimental one, than it was the case of form II. These similarities concern primarily conformation of a molecule found inside the crystal and/or similar hydrogen bonding interactions. However, serendipitous agreement in terms of $^1$H only or $^{13}$C only RMS data, but not in terms of both, was also observed for this form.

***Comparison of Z' = 1 (form II) and Z' = 2 (form I) polymorphs.***

DSC measurements for both neat polymorphs of furazidin indicate that they have similar thermal stability, which at the same time is a measure of their enthalpy of formation. Therefore, we can expect that their total energies obtained from quantum chemical calculations will also be similar. This indeed is the case. In the force field crystal energy landscape crystal structures identified as forms I and II are separated by 4 kJ/mol, with form II having lower total energy than form I, but after the DFT optimization this trend was reversed, and form I was found to be more energetically stable than form II, with total DFT energy 1.6 kJ/mol lower. Regardless of the level of theory used, the observed differences in energy are small and support the existence of these two polymorphic forms. Table 1 summarizes changes to the unit cell parameters in the crystal structures of furazidin form I and II obtained from force field (CSP) calculations, after DFT-D2 geometry optimization, as well as after Rietveld refinement of the computationally generated structures against respective powder X-Ray diffraction patterns. As can be seen optimization of the CSP-generated structures at higher level of theory (DFT-D2) did not change unit cell parameters significantly, with Rietveld refinement having more noticeable impact on the regarded parameters. All cif files corresponding to the described crystal structures are attached to this work as Supporting Information Files.

**Table 1.** Comparison of unit cell parameters for crystal structures of furazidin forms I and II obtained after force field and DFT-D2 geometry optimization, as well as after Rietveld refinement. The obtained cif files are attached to this work as Supporting Information either in their native cif format (for Rietveld refined structures) or as plain text files (for other cif files).

| Unit cell parameters | Force field optimized structure | DFT-D2 optimized structure | Rietveld refined structure |
|---|---|---|---|
| form I | | | |
| a | 22.388 | 22.440 | 22.183(22) |
| b | 12.654 | 12.655 | 12.546(7) |
| c | 5.334 | 5.334 | 5.3092(16) |
| α | 50.937 | 50.957 | 51.074(3) |
| β | 92.532 | 92.507 | 92.16(3) |
| γ | 97.159 | 97.022 | 97.31(11) |
| form II | | | |
| a | 11.659 | 11.898 | 11.670(7) |
| b | 11.944 | 11.636 | 11.4666(12) |
| c | 8.466 | 8.467 | 8.494(4) |
| β | 107.22 | 106.65 | 106.848(14) |

The lowest energy region of the force field crystal energy landscape for Z' = 1 and Z' = 2 crystal forms (first 10 kJ/mol above the global minimum, Figure 11d) contains significantly larger amount of Z' = 2 structures. This is probably the result of a larger number of possible packing arrangements, but it may also originate from a better energetic compensation offered by a possibility of formation of Z' = 2 structures, as pointed out before. A comparison of form I and form II structures (Figures 11a and 11b) indicates that form II is visibly more densely packed (its density after DFT-D2 optimization is equal to 1.563 g/cm$^3$), in comparison to form I (d = 1.506 g/cm$^3$), and yet this latter form has lower DFT-D2 energy. On the other hand, conformer found in this form has relative gas phase energy 11.94 kJ/mol higher than the global minimum, which is only 3 kJ/mol higher than is the case of form II (for an overlay of furazidin molecules see Figure 11c). This means that the intermolecular energy contribution to the total energy of form I is 4.6 kJ/mol higher than that of form II. This in part support the conclusion that Z' = 2 structure offers furazidin molecules more favorable intermolecular interactions and therefore a better energetic compensation. The differences in molecular environment of a single furazidin molecules in both polymorphs are highlighted by Hirshfeld surfaces shown in Figures S9 (Supporting Information), while the calculated energy frameworks showing the contribution of dispersion and electrostatic energies to the total energies of the crystals are shown in Table S7, Supporting Information. Although both polymorphs of furazidin are structurally distinct, their global interaction topologies are apparently comparable as evident from energy framework analysis (see Tables S3-S7, Supporting Information).

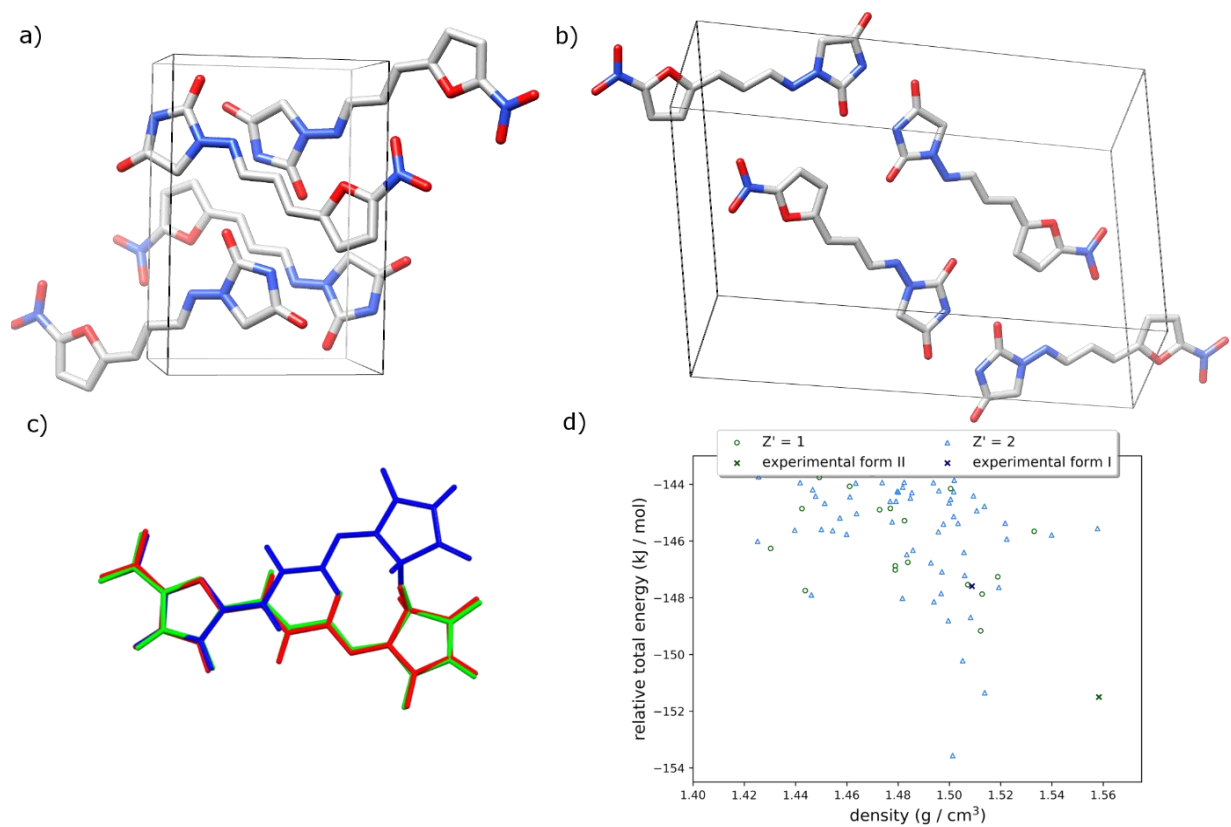

**Figure 11.** Packing arrangement for the experimental crystal structures of form II (a) and form I (b) of furazidin, an overlay of furazidin molecules present in form I (red and green) and in form II (blue) (c), and the comparison of the lowest energy region of crystal energy landscapes of Z' = 1 and Z' = 2 structures of furazidin (d).

Finally, we would like to point to similarities of supramolecular synthons present in two polymorphs of furazidin with the synthons observed in two crystal structures of a similar molecule. As mentioned before, in the CCDC database no crystal structures of neat furazidin has been published so far. There are however crystal structures of nitrofurantion, which differs from furazidin only in the length of a carbon chain linking tetrahydrofuran and hydantion rings. Nitrofurantoin, similarly like furazidin, has two polymorphic forms, a triclinic one and a monoclinic one. These polymorphs, designated as α and β forms, crystallize in *P*-1 and *P*2$_1$/n space groups, respectively (the respective CCDC refcodes for these structures are LABJON01 and LABJON), (Bertolasi *et al.*, 1993; Pienaar *et al.*, 1993), both with one nitrofurantoin molecule in an asymmetric part of a unit cell. The first of these polymorphs, the triclinic one, display very similar intermolecular interactions to the triclinic form of furazidin (form I), with dimers interacting *via* N-H...O from hydantoin rings (Figure S11a, Supporting Information). In contrast, in the β polymorph there are only some supramolecular synthons which are the same as in form II of furazidin, namely N-H...O

(carbonyl group) interactions, with no N-H...O ($NO_2$ group) hydrogen bonds. The main difference in the intermolecular interactions between the two structures lies in the orientation of the regarded molecules: in monoclinic polymorph of nitrofurantion the molecules are arranged in a head-to-head manner, where the 'head' is a hydantoin ring (Figure S11b, Supporting Information), while in furazidin monoclinic polymorph there is a head-to-tail arrangement, *i.e.* $NO_2$ group substituted to the furan ring is directed towards the hydantion ring from the second molecule.

**Conclusions**

In this work two previously uncharacterized polymorphs of furazidin were described using NMR crystallography and crystal structure prediction calculations. The described Z' = 1 and Z' = 2 polymorphic forms display similar thermal and energetic stability and were found to crystallize in $P2_1/c$ and P-1 space groups, respectively. Their structural determination was possible only after combining data from very fast magic angle spinning solid-state NMR experiments, powder X-ray diffraction and crystal structure prediction calculations. In terms of NMR crystallography application to the structural determination of microcrystalline substances, our results underline the need for caution when analyzing the agreement between theoretical and experimental NMR data, indicating the cases of false-positive matches, especially when accounting for NMR data obtained only for one nuclei (*i.e.* only $^1$H or $^{13}$C). On the other hand, a case in which the NMR data for both nuclei are serendipitously in very good agreement with the experiment has not been observed. It is therefore advisable always to look at both $^1$H and $^{13}$C data, when elucidating crystal structure using NMR crystallography. We feel that there is still not enough data in the literature identifying and analyzing cases of false-positive matches in terms of solid state NMR data and/or dealing with the question of how good agreement in terms of NMR data is good enough for different crystal systems (Z' > 1 polymorphs, hydrates, solvates, flexible molecules). Such studies can be a valuable source of information as to what can be expected when using NMR crystallography in crystal structure determination.

As a summary, we would like to stress that crystal structure determination from a microcrystalline powder using CSP-NMR approach is much more difficult, when dealing with Z' > 1 polymorphs. This concerns almost all stages of the structure determination process, starting from the necessity of accounting for more degrees of freedom in the CSP search (both conformational and in terms of molecules rotations and translations in a crystal), through difficulties in extracting viable structural constrains from the NMR experiments (mainly due to the signals overlap), to reliable identification of a correct structure from among the numerous ones generated in the CSP process (mostly due to a larger amount of structures giving reasonably good agreement with the NMR experiment).

**Acknowledgements**

This work was financially supported by Polish National Science Center under Sonata 14 grant No. UMO-2018/31/D/ST4/01995. The Polish Infrastructure for Supporting Computational Science in the European Research Space (PL-GRID) is gratefully acknowledged for providing computational resources. This project has received funding from the European Union's Horizon 2020 research and innovation program under grant agreement No E190200237 (EUSMI). The Authors are very grateful to Wojciech Dudek for his help in the adaptation of the CSPy code to Polish supercomputing infrastructure.

# Supporting Information

to

# Crystal structures of two furazidin polymorphs revealed by a joint effort of crystal structure prediction and NMR crystallography

Marta K. Dudek,[1]* Piotr Paluch,[1] Edyta Pindelska[2]*

[1]Centre of Molecular and Macromolecular Studies, Polish Academy of Sciences, Sienkiewicza 112, 90363 Lodz, Poland, e-mail: mdudek@cbmm.lodz.pl

[2]Faculty of Pharmacy, Medical University of Warsaw, Banacha 1, 02097 Warsaw, Poland, e-mail: edyta.pindelska@wum.edu.pl

## 1. FT-IR spectra of furazidin form I and II

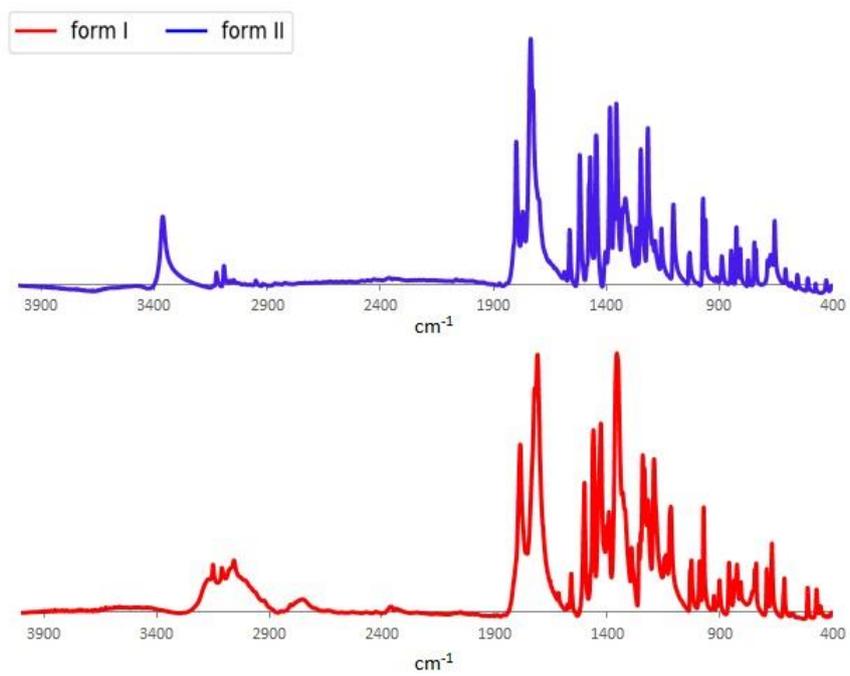

**Figure S1.** FT-IR spectra of furazidin form I and II.

## 2. DSC thermograms of furazidin form I and II

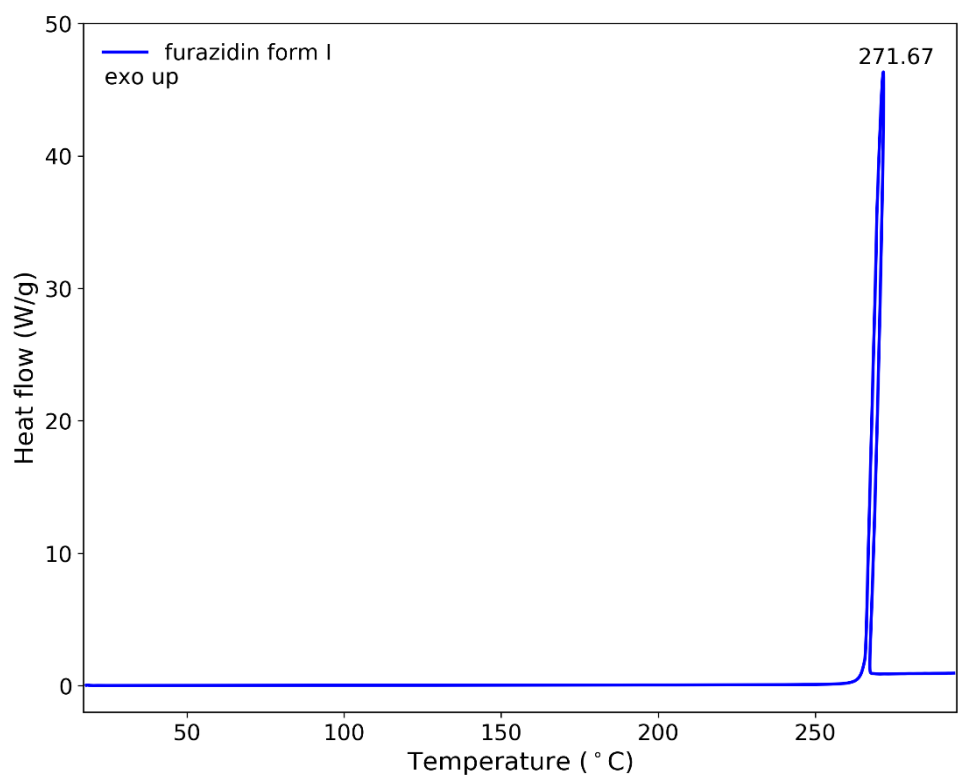

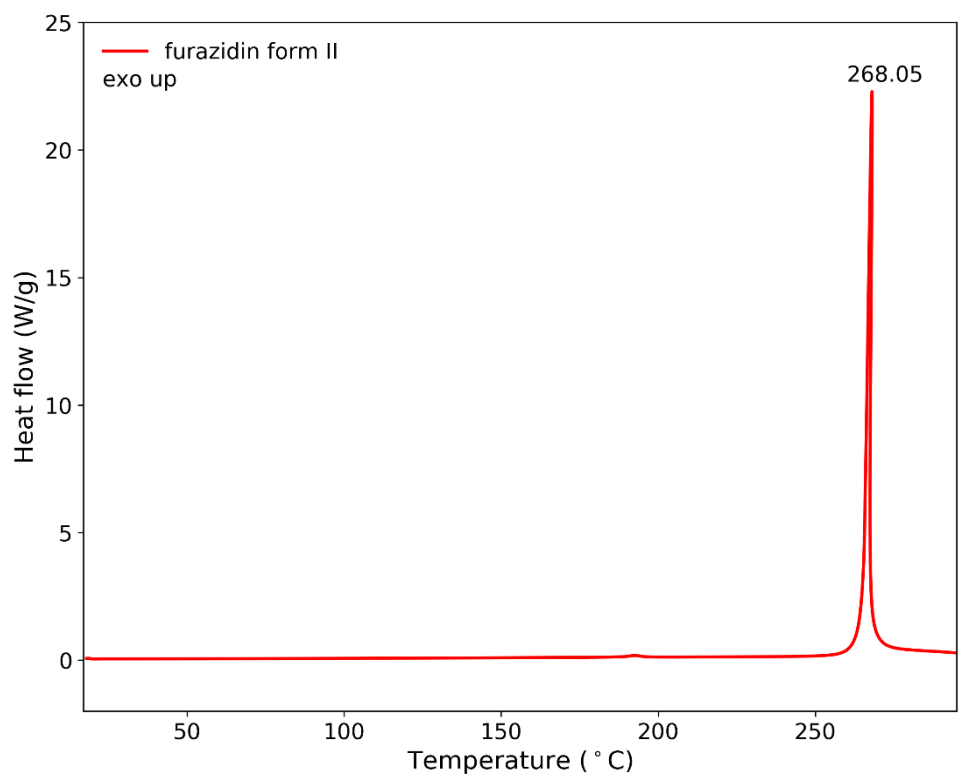

**Figure S2.** DSC for furazidin form I and II at scan rate 2 K/min. The shown temperatures correspond to the melting/decomposition points, while the respective enthalpies of fusion for the observed exothermic events are equal to 1231 and 1205 J/g. Note, however, that due to a significant amount of heat emitted during these events the given values may be burdened with error. A small

exothermic event at 192 °C is associated with a weight loss of 0.23% of the sample of furazidin form II (see TGA plot below, registered in the temperature range of 20 – 220 °C) and is not reversible, nor visible in the second run of heating, therefore it is probably associated with a decomposition of a small amount of impurities present in the sample. The performed experiments indicate that forms I and II are enantiotropic polymorphs.

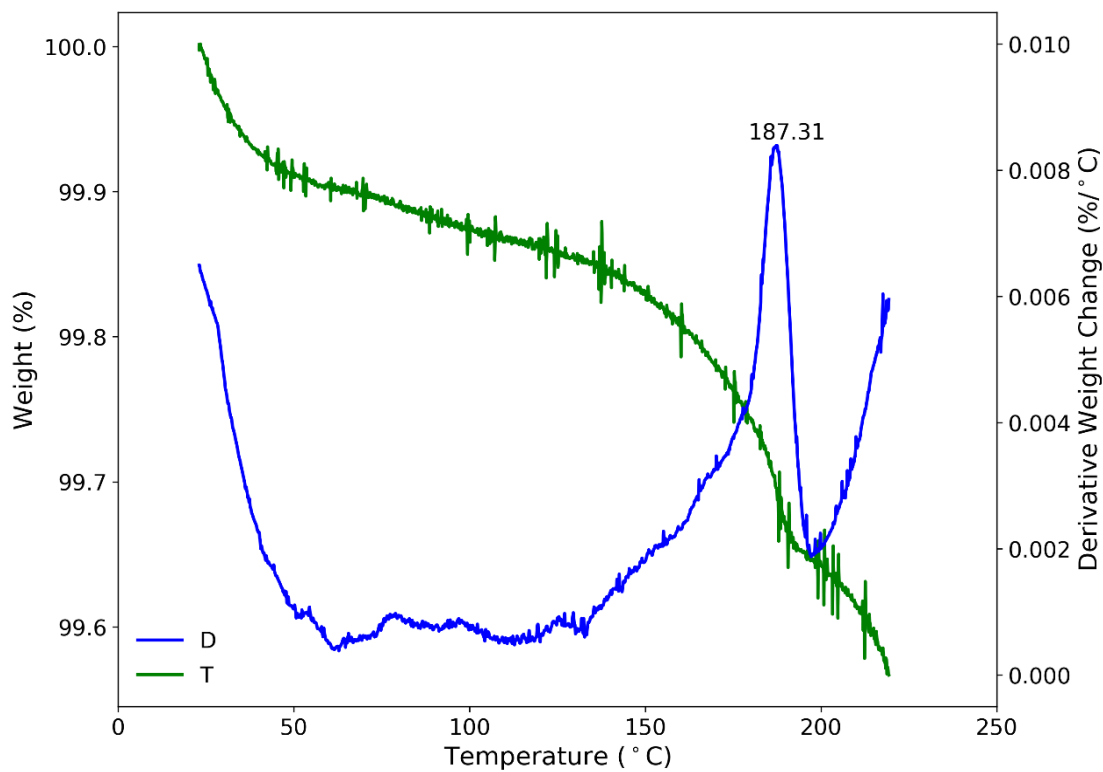

**Figure S3.** TGA and DTA plots registered for furazidin form II in the temperature range of 20 – 220 °C showing 0.23% weight loss associated with an exothermic event observed in the DSC plot at 192 °C.

## 3. Additional solid-state NMR spectra registered for form I and II of furazidin

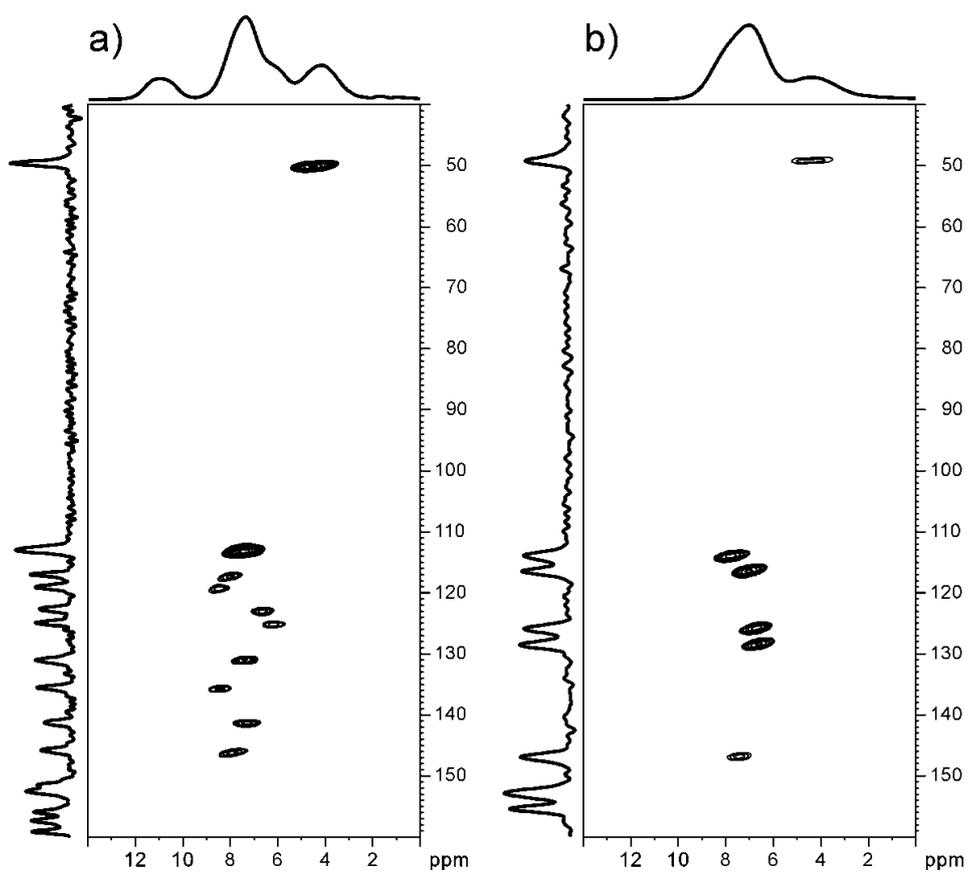

**Figure S4.** *inv*-$^1$H-$^{13}$C HETCOR spectrum registered with short second contact time (50 μs) for furazidin form I (a) and (II); spinning speed = 60 kHz; spectrum a) was registered at an 800 MHz spectrometer to obtain good separation of correlation peaks, whereas spectrum b) was registered with a 600 MHz spectrometer.

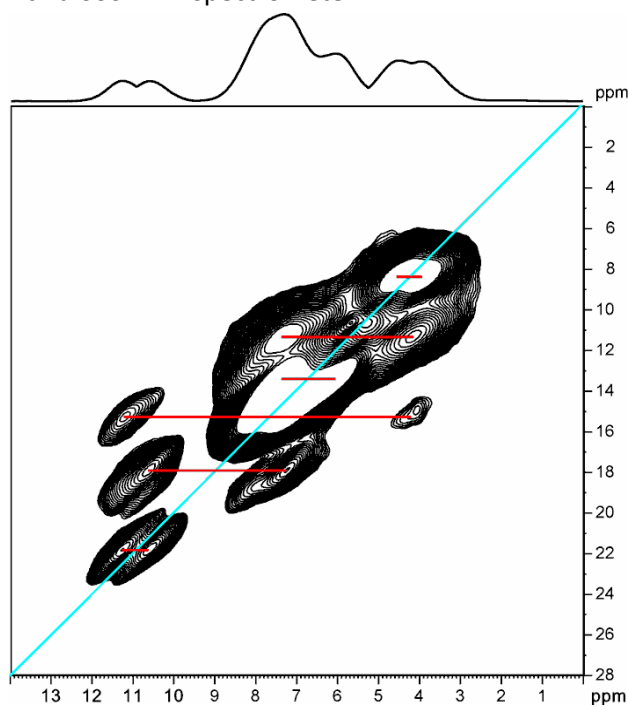

**Figure S5.** $^1$H-$^1$H Back-to-Back correlation spectrum for furazidin form I; spinning speed = 60 kHz.

## 4. Additional CSP crystal energy landscapes for Z' = 2 structure of furazidin

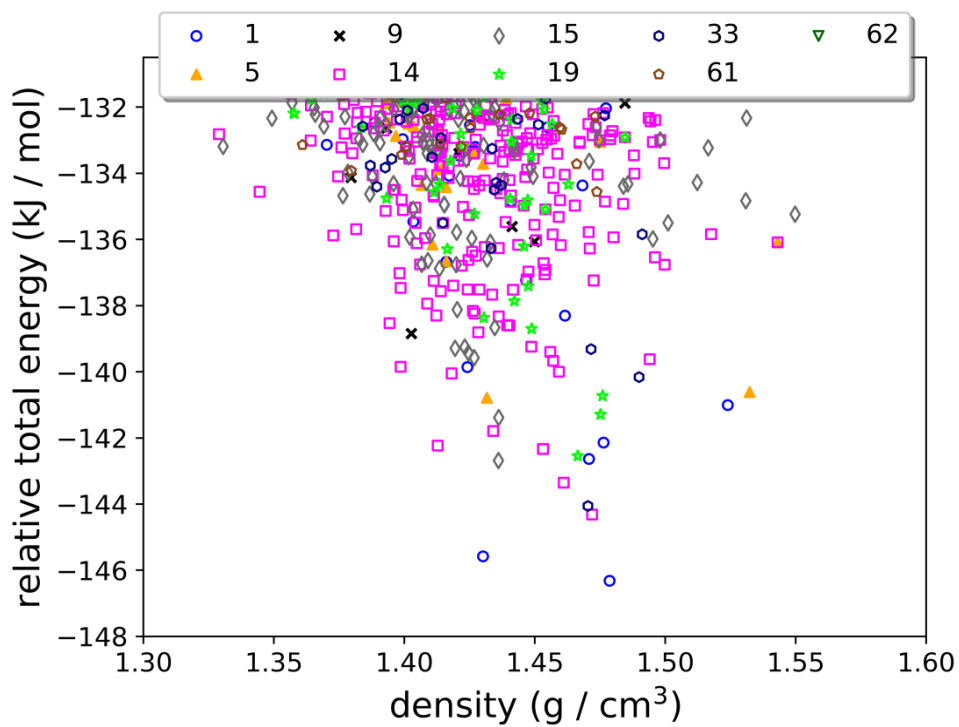

**Figure S6.** Crystal energy landscape obtained after CSP search for the pair of conformer indicated by comparison with NMR experiments in space groups P1, C2, Cc, P2$_1$/c, C2/c, P2$_1$2$_1$2$_1$, Pna2$_1$, Pbca and Pnma

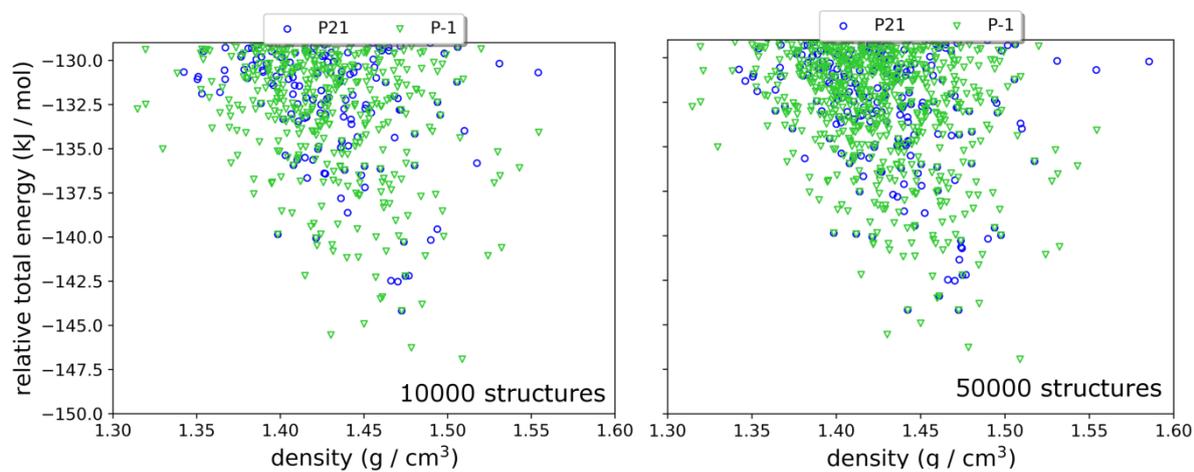

**Figure S7.** Comparison of CSP landscapes for the selected pair of conformers tested in P2$_1$ and P1 space groups, in which 10000 or 50000 valid crystal structures were generated.

## 5. The assignment of ¹H and ¹³C solid-state NMR chemical shifts for forms I and II of furazidin

Table S1. ¹H and ¹³C solid-state NMR chemical shifts of furazidin form II

| atom | δ(¹³C) | δ(¹H) |
|---|---|---|
| C2 | 155.4 | |
| C3 | 113.9 | 7.8 |
| C4 | 116.3 | 7.0 |
| C5 | 152.8 | |
| Cα | 125.7 | 6.75 |
| Cβ | 128.4 | 6.6 |
| Cγ | 146.9 | 7.4 |
| C2' | 152.8 | |
| C4' | 167.2 | |
| C5' | 48.8 | 4.7; 4.0 |
| NH | | 8.1 |

Table S2. Two variants of the assignment of ¹H and ¹³C solid-state NMR chemical shifts of furazidin form I

| | | variant I | | variant II | |
|---|---|---|---|---|---|
| molecule | atom | δ(¹³C) | δ(¹H) | δ(¹³C) | δ(¹H) |
| A | C2 | 157.4 | | 159.2 | |
| | C3 | 113.1 | 7.2 | 113.1 | 7.0 |
| | C4 | 119.1 | 7.8 | 119.1 | 7.8 |
| | C5 | 151.8 | | 151.8 | |
| | Calpha | 125.0 | 5.8 | 122.8 | 6.4 |
| | Cbeta | 131.2 | 7.3 | 135.5 | 8.2 |
| | Cgamma | 141.3 | 6.95 | 141.3 | 6.95 |
| | C2' | 152.8 | | 152.8 | |
| | C4' | 173.6 | | 173.6 | |
| | C5' | 49.6 | 4.5; 4.2 | 49.6 | 4.5; 4.2 |
| | NH | | 10.8 | | 10.8 |
| B | C2 | 159.2 | | 157.4 | |
| | C3 | 113.1 | 7.0 | 113.1 | 7.2 |
| | C4 | 117.1 | 7.6 | 117.1 | 7.6 |
| | C5 | 151.8 | | 151.8 | |
| | Calpha | 122.8 | 6.4 | 125.0 | 5.8 |
| | Cbeta | 135.5 | 8.2 | 131.2 | 7.3 |
| | Cgamma | 145.9 | 7.3 | 145.9 | 7.3 |
| | C2' | 155.9 | | 155.9 | |
| | C4' | 167.3 | | 167.3 | |
| | C5' | 49.6 | 4.1; 3.9 | 49.6 | 4.1; 3.9 |
| | NH | | 11.3 | | 11.3 |

## 6. Comparison of simulated and experimental PXRD diffractograms for form I and II of furazidin

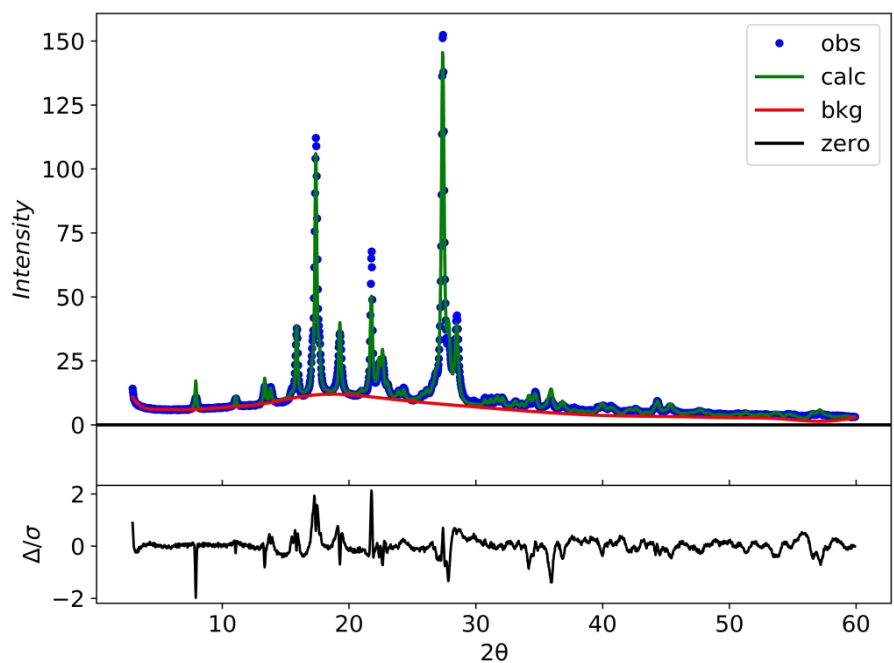

**Figure S8.** Overlay of the experimental and simulated PXRD diffractogram for furazidin form II.

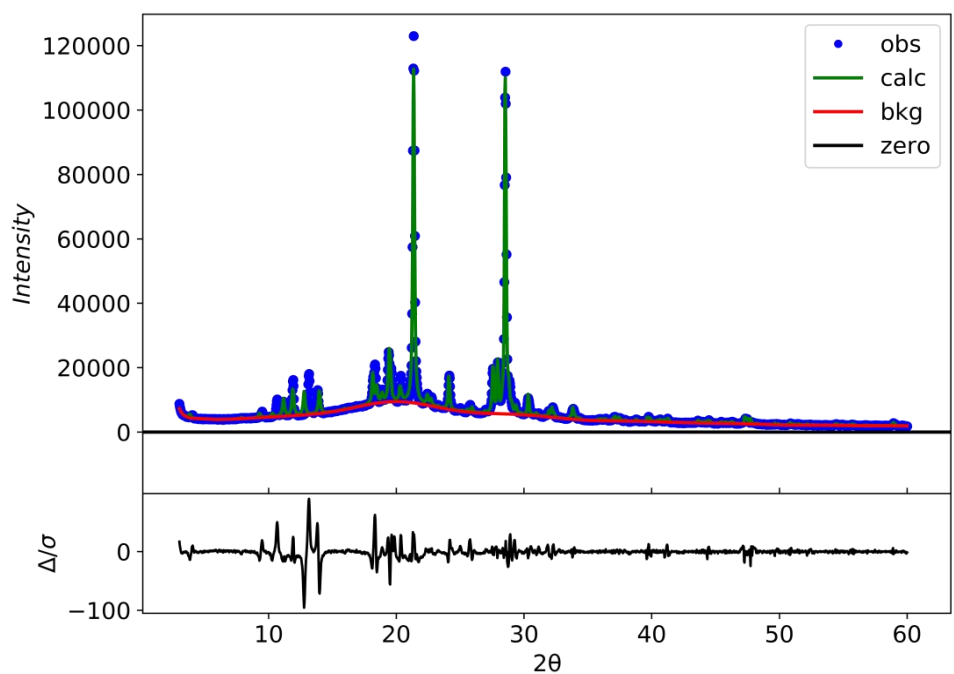

**Figure S9.** Overlay of the experimental and simulated PXRD diffractogram for furazidin form I.

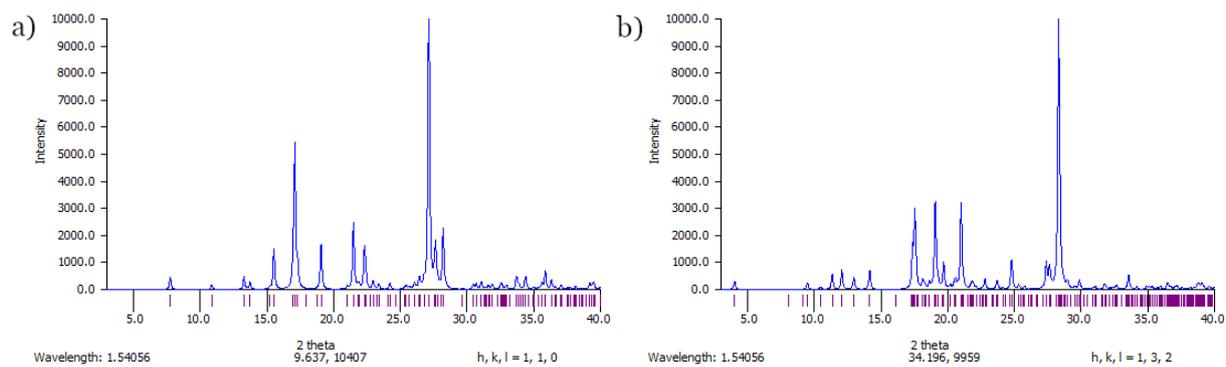

**Figure S10.** Simulated PXRD patterns for computationally generated structures of form II (a) and I (b).

## 7. Hirshfeld surfaces and energy frameworks for calculated for crystal structures of form I and II of furazidin

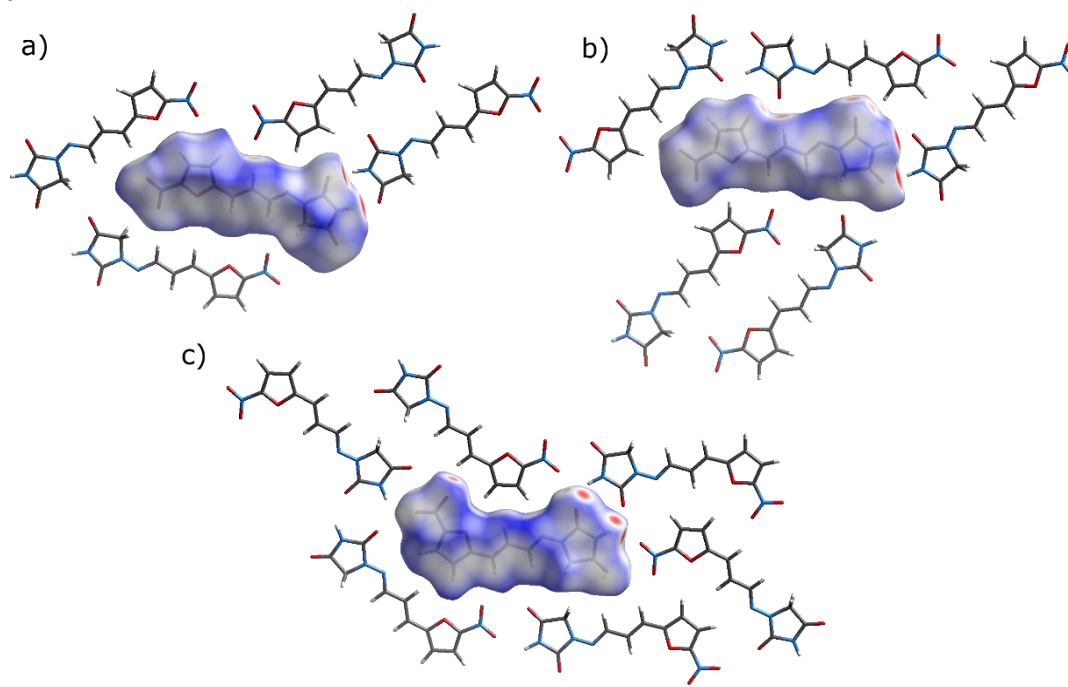

**Figure S11.** The HSs mapped with d$_{norm}$ for furazidin molecules: a) form I, molecule A, b) form I, molecule B, and c) form II

**Table S3.** The interaction energies (kJ/mol) obtained from energy framework calculation for furazidin form I molecule A.

| | N | Symop | R | Electron Density | E_ele | E_pol | E_dis | E_rep | E_tot |
|---|---|---|---|---|---|---|---|---|---|
| | 0 | x, y, z | 5.33 | B3LYP/6-31G(d,p) | -4.9 | -5.4 | -33.2 | 18.1 | -26.9 |
| | 0 | -x, -y, -z | 7.36 | B3LYP/6-31G(d,p) | -25.7 | -9.5 | -18.2 | 35.7 | -27.9 |
| | 0 | -x, -y, -z | 7.74 | B3LYP/6-31G(d,p) | -17.4 | -3.7 | -30.3 | 12.6 | -39.8 |
| | 1 | - | 9.81 | B3LYP/6-31G(d,p) | -2.6 | -0.1 | -0.6 | 0.0 | -3.4 |
| | 1 | - | 13.67 | B3LYP/6-31G(d,p) | -4.9 | -5.4 | -33.2 | 18.1 | -26.9 |
| | 1 | - | 7.58 | B3LYP/6-31G(d,p) | -25.7 | -9.5 | -18.2 | 35.7 | -27.9 |
| | 0 | -x, -y, -z | 12.54 | B3LYP/6-31G(d,p) | -12.1 | -2.2 | -11.4 | 5.9 | -20.7 |
| | 0 | -x, -y, -z | 15.75 | B3LYP/6-31G(d,p) | 4.1 | -0.2 | -1.0 | 0.1 | 3.4 |
| | 2 | x, y, z | 5.33 | B3LYP/6-31G(d,p) | -4.9 | -5.4 | -33.2 | 18.1 | -26.9 |

**Table S4.** The interaction energies (kJ/mol) obtained from energy framework calculation for furazidin form I molecule B.

| | N | Symop | R | Electron Density | E_ele | E_pol | E_dis | E_rep | E_tot |
|---|---|---|---|---|---|---|---|---|---|
| | 2 | x, y, z | 5.33 | B3LYP/6-31G(d,p) | -4.9 | -5.4 | -33.2 | 18.1 | -26.9 |
| | 1 | -x, -y, -z | 7.36 | B3LYP/6-31G(d,p) | -25.7 | -9.5 | -18.2 | 35.7 | -27.9 |
| | 1 | -x, -y, -z | 7.74 | B3LYP/6-31G(d,p) | -17.4 | -3.7 | -30.3 | 12.6 | -39.8 |
| | 1 | - | 9.81 | B3LYP/6-31G(d,p) | -2.6 | -0.1 | -0.6 | 0.0 | -3.4 |
| | 1 | - | 13.67 | B3LYP/6-31G(d,p) | -4.9 | -5.4 | -33.2 | 18.1 | -26.9 |
| | 1 | - | 7.58 | B3LYP/6-31G(d,p) | -25.7 | -9.5 | -18.2 | 35.7 | -27.9 |
| | 1 | -x, -y, -z | 12.54 | B3LYP/6-31G(d,p) | -12.1 | -2.2 | -11.4 | 5.9 | -20.7 |
| | 1 | -x, -y, -z | 15.75 | B3LYP/6-31G(d,p) | 4.1 | -0.2 | -1.0 | 0.1 | 3.4 |

**Table S5.** The interaction energies (kJ/mol) obtained from energy framework calculation for furazidin form II.

| | N | Symop | R [Å] | Electron Density | E_ele | E_pol | E_dis | E_rep | E_tot |
|---|---|---|---|---|---|---|---|---|---|
| | 1 | -x, -y, -z | 10.01 | B3LYP/6-31G(d,p) | -18.4 | -7.0 | -20.6 | 18.4 | -31.2 |
| | 1 | -x, -y, -z | 5.76 | B3LYP/6-31G(d,p) | -4.7 | -3.7 | -35.7 | 18.1 | -27.6 |
| | 2 | -x, y+1/2, -z+1/2 | 9.31 | B3LYP/6-31G(d,p) | -7.5 | -3.0 | -9.4 | 5.9 | -14.7 |
| | 2 | -x, y+1/2, -z+1/2 | 7.91 | B3LYP/6-31G(d,p) | -15.4 | -6.0 | -23.2 | 30.2 | -22.2 |
| | 1 | -x, -y, -z | 9.20 | B3LYP/6-31G(d,p) | -20.6 | -2.5 | -38.2 | 28.3 | -39.4 |
| | 2 | x, -y+1/2, z+1/2 | 4.83 | B3LYP/6-31G(d,p) | -11.8 | -6.0 | -38.3 | 17.6 | -39.4 |
| | 1 | -x, -y, -z | 12.41 | B3LYP/6-31G(d,p) | -30.2 | -4.6 | -9.3 | 10.2 | -37.2 |
| | 2 | x, y, z | 12.47 | B3LYP/6-31G(d,p) | -13.4 | -2.3 | -12.8 | 8.5 | -21.8 |
| | 1 | x, -y+1/2, z+1/2 | 14.89 | B3LYP/6-31G(d,p) | -7.0 | -2.7 | -5.8 | 6.0 | -10.7 |

**Table S6.** Scale factors for benchmarked energy models (see: Mackenzie, C. F., Spackman, P. R., Jayatilaka, D. & Spackman, M. A. (2017). IUCrJ 4, 575-587).

| Energy Model | k_ele | k_pol | k_disp | k_rep |
|---|---|---|---|---|
| CE-HF ... HF/3-21G electron densities | 1.019 | 0.651 | 0.901 | 0.811 |
| CE-B3LYP ... B3LYP/6-31G(d,p) electron densities | 1.057 | 0.740 | 0.871 | 0.618 |

**Table 7.** Energy frameworks of total energy (blue) along with decomposed electrostatic (red) and dispersive (green) components for the crystal structures of furazidin. Energies with a magnitude less than 20 kJ mol$^{-1}$ have been omitted.

| Energy | form I_ molecule A | form I_ molecule B | form_II |
|---|---|---|---|
| electrostatic | 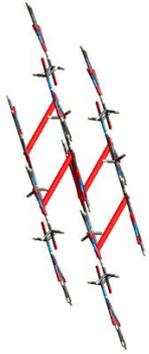 | 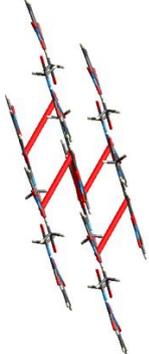 | 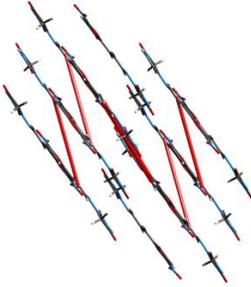 |
| dispersion | 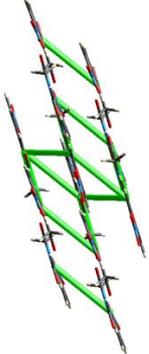 | 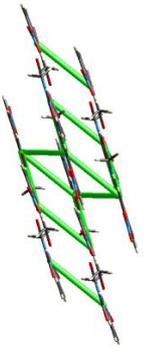 | 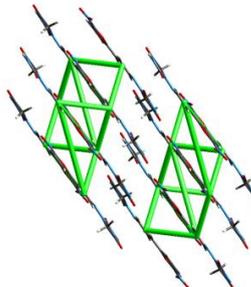 |
| total | 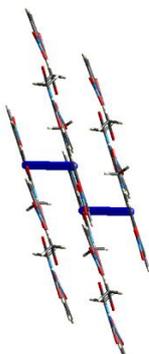 | 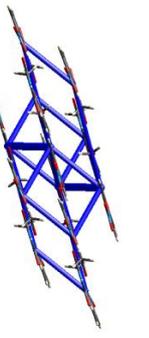 | 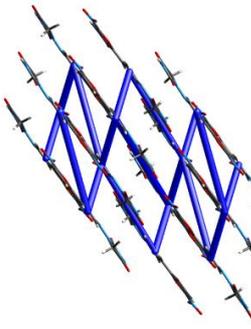 |

## 8. The $R^2$, RMS, slope and intercept values of all model crystal structures re-optimized at DFT-D2 level of theory for forms I and II of furazidin

**Table S3.** DFT-D2 relative energy and linear regression data obtained after comparison of the calculated and experimental $^1$H and $^{13}$C NMR data for furazidin form II.

| rank | relative energy (kJ/mol) | $^1$H NMR data | | | | $^{13}$C NMR data | | | |
|---|---|---|---|---|---|---|---|---|---|
| | | $R^2$ | intercept | slope | RMS (ppm) | $R^2$ | intercept | slope | RMS (ppm) |
| 1 | 0.00 | 0.7228 | -1.594 | 33.75 | 0.84 | 0.9854 | -1.008 | 169.41 | 3.92 |
| 2 | 0.03 | 0.6813 | -1.323 | 31.82 | 0.93 | 0.9879 | -1.019 | 170.68 | 3.57 |
| 3 | 1.73 | 0.9551 | -1.085 | 30.83 | 0.30 | 0.9938 | -1.002 | 169.34 | 2.55 |
| 4 | 6.34 | 0.8129 | -1.107 | 30.75 | 0.65 | 0.9702 | -1.006 | 170.68 | 5.65 |
| 5 | 6.65 | 0.7612 | -1.137 | 31.12 | 0.76 | 0.9809 | -0.981 | 165.65 | 4.50 |
| 6 | 14.13 | 0.8380 | -1.119 | 30.55 | 0.60 | 0.9763 | -1.009 | 171.01 | 5.03 |
| 7 | 15.94 | 0.8203 | -1.017 | 30.25 | 0.64 | 0.9729 | -1.013 | 172.31 | 5.38 |
| 8 | 18.17 | 0.7110 | -1.464 | 32.41 | 0.87 | 0.9742 | -1.023 | 172.62 | 5.25 |
| 9 | 18.21 | 0.8777 | -1.174 | 31.01 | 0.51 | 0.9854 | -0.988 | 167.04 | 3.93 |
| 10 | 18.61 | 0.9081 | -1.014 | 30.34 | 0.43 | 0.9697 | -1.000 | 171.34 | 5.70 |
| 11 | 19.65 | 0.7365 | -1.505 | 33.58 | 0.82 | 0.9834 | -1.008 | 170.76 | 4.20 |
| 12 | 22.22 | 0.9025 | -1.195 | 31.37 | 0.45 | 0.9600 | -0.997 | 169.72 | 6.58 |
| 13 | 25.10 | 0.6617 | -1.448 | 32.98 | 0.97 | 0.9740 | -0.993 | 169.79 | 5.27 |
| 14 | 30.08 | 0.7716 | -1.610 | 33.77 | 0.74 | 0.9709 | -1.019 | 173.18 | 5.59 |
| 15 | 31.81 | 0.6798 | -0.779 | 28.87 | 0.94 | 0.9912 | -1.010 | 170.22 | 3.03 |
| 16 | 33.80 | 0.9240 | -1.199 | 31.61 | 0.39 | 0.9680 | -1.001 | 171.23 | 5.86 |

**Table S4.** DFT-D2 relative energy and linear regression data obtained after comparison of the calculated and experimental (variant II) $^1$H and $^{13}$C NMR data for furazidin form I.

| rank | relative energy (kJ/mol) | $^1$H NMR data | | | | $^{13}$C NMR data | | | |
|---|---|---|---|---|---|---|---|---|---|
| | | $R^2$ | intercept | slope | RMS (ppm) | $R^2$ | intercept | slope | RMS (ppm) |
| 1 | 0.00 | 0.9633 | -1.173 | 31.13 | 0.41 | 0.9944 | -1.019 | 171.30 | 2.45 |
| 2 | 2.76 | 0.9249 | -1.082 | 30.46 | 0.59 | 0.9892 | -1.021 | 171.43 | 3.41 |
| 3 | 5.65 | 0.9452 | -1.02 | 30.14 | 0.50 | 0.9943 | -1.016 | 171.22 | 2.46 |
| 4 | 10.02 | 0.824 | -0.974 | 29.65 | 0.96 | 0.9917 | -0.981 | 164.90 | 2.99 |
| 5 | 10.55 | 0.9297 | -1.059 | 30.41 | 0.57 | 0.9806 | -1.017 | 171.96 | 4.58 |
| 6 | 10.91 | 0.802 | -0.972 | 29.85 | 1.03 | 0.9820 | -0.977 | 166.09 | 4.41 |
| 7 | 11.55 | 0.8651 | -0.746 | 28.46 | 0.82 | 0.9756 | -0.998 | 170.28 | 5.15 |
| 8 | 12.24 | 0.9451 | -1.152 | 31.19 | 0.50 | 0.9856 | -1.010 | 171.57 | 3.95 |
| 9 | 13.18 | 0.8971 | -1.054 | 30.21 | 0.70 | 0.9820 | -0.978 | 166.03 | 4.42 |
| 10 | 13.42 | 0.8025 | -0.788 | 29.18 | 1.03 | 0.9916 | -0.985 | 166.41 | 3.00 |
| 11 | 13.46 | 0.7995 | -0.932 | 29.85 | 1.04 | 0.9783 | -0.980 | 165.85 | 4.86 |
| 12 | 14.07 | 0.8653 | -0.851 | 29.29 | 0.82 | 0.9875 | -0.989 | 166.69 | 3.67 |
| 13 | 14.34 | 0.8754 | -0.978 | 30.31 | 0.79 | 0.9835 | -0.952 | 163.07 | 4.22 |
| 14 | 14.68 | 0.9278 | -0.972 | 29.90 | 0.58 | 0.9677 | -0.998 | 170.67 | 5.96 |
| 15 | 15.06 | 0.8708 | -1.133 | 30.80 | 0.80 | 0.9859 | -0.990 | 167.60 | 3.90 |
| 16 | 16.18 | 0.7923 | -0.897 | 29.37 | 1.07 | 0.9874 | -0.992 | 167.14 | 3.68 |
| 17 | 16.61 | 0.7758 | -0.685 | 28.05 | 1.12 | 0.9873 | -0.999 | 170.42 | 3.70 |
| 18 | 17.87 | 0.7822 | -1.101 | 30.87 | 1.10 | 0.9887 | -0.972 | 165.26 | 3.49 |
| 19 | 18.07 | 0.8524 | -0.959 | 30.15 | 0.87 | 0.9826 | -0.998 | 169.74 | 4.34 |
| 20 | 18.39 | 0.8202 | -0.791 | 28.96 | 0.97 | 0.9750 | -0.985 | 168.10 | 5.22 |
| 21 | 18.70 | 0.8668 | -1.031 | 30.22 | 0.82 | 0.9850 | -0.996 | 169.65 | 4.03 |
| 22 | 19.92 | 0.8468 | -1.057 | 30.43 | 0.89 | 0.9887 | -0.997 | 167.76 | 3.48 |
| 23 | 19.99 | 0.8934 | -0.949 | 30.09 | 0.72 | 0.9743 | -0.975 | 166.87 | 5.30 |
| 24 | 20.00 | 0.8305 | -0.951 | 29.86 | 0.94 | 0.9800 | -0.986 | 167.41 | 4.66 |
| 25 | 20.17 | 0.8042 | -0.912 | 29.51 | 1.03 | 0.9910 | -1.004 | 169.19 | 3.11 |
| 26 | 20.23 | 0.8389 | -1.006 | 30.08 | 0.91 | 0.9787 | -0.984 | 167.39 | 4.81 |
| 27 | 20.27 | 0.7680 | -1.151 | 30.90 | 1.14 | 0.9913 | -0.965 | 162.66 | 3.05 |
| 28 | 20.80 | 0.7908 | -0.958 | 29.83 | 1.07 | 0.9880 | -0.975 | 163.73 | 3.59 |
| 29 | 21.07 | 0.8988 | -1.014 | 30.10 | 0.70 | 0.9883 | -0.996 | 168.88 | 3.55 |
| 30 | 21.39 | 0.6847 | -0.788 | 28.29 | 1.41 | 0.9814 | -0.962 | 162.78 | 4.49 |
| 31 | 21.63 | 0.8070 | -1.063 | 30.24 | 1.02 | 0.9949 | -0.954 | 162.02 | 2.32 |
| 32 | 21.95 | 0.7480 | -0.915 | 29.31 | 1.21 | 0.9851 | -0.973 | 164.94 | 4.01 |
| 33 | 22.08 | 0.7279 | -0.833 | 29.27 | 1.27 | 0.9818 | -0.972 | 165.56 | 4.44 |
| 34 | 22.84 | 0.7638 | -0.919 | 29.59 | 1.16 | 0.9896 | -0.945 | 160.20 | 3.34 |
| 35 | 24.70 | 0.8577 | -0.813 | 29.41 | 0.85 | 0.9861 | -0.975 | 166.78 | 3.88 |
| 36 | 24.82 | 0.8410 | -1.081 | 30.45 | 0.91 | 0.9799 | -0.988 | 167.84 | 4.67 |
| 37 | 24.91 | 0.7467 | -0.975 | 30.00 | 1.21 | 0.9833 | -0.968 | 164.71 | 4.25 |
| 38 | 25.02 | 0.8233 | -1.083 | 30.92 | 0.96 | 0.9835 | -0.994 | 169.05 | 4.22 |
| 39 | 25.08 | 0.6689 | -0.836 | 29.38 | 1.46 | 0.9788 | -0.934 | 160.30 | 4.80 |
| 40 | 27.74 | 0.7732 | -0.846 | 28.95 | 1.13 | 0.9835 | -0.967 | 164.66 | 4.22 |
| 41 | 28.91 | 0.7546 | -1.173 | 30.97 | 1.19 | 0.9811 | -0.971 | 163.92 | 4.52 |
| 42 | 29.66 | 0.7329 | -1.169 | 31.36 | 1.26 | 0.9868 | -0.958 | 162.99 | 3.78 |

**Table S5.** DFT-D2 relative energy and linear regression data obtained after comparison of the calculated and experimental (variant II) $^1$H and $^{13}$C NMR data for furazidin form I. These data are for structures obtained from an additional CSP search for the best conformer indicated by the data from Table S4 tested in the space groups given in this table.

| space group | relative energy (kJ/mol) | $^1$H NMR data | | | | $^{13}$C NMR data | | | |
|---|---|---|---|---|---|---|---|---|---|
| | | $R^2$ | intercept | slope | RMS (ppm) | $R^2$ | intercept | slope | RMS (ppm) |
| 14 | 17.67 | 0.9195 | -0.866 | 29.42 | 0.62 | 0.9909 | -1.018 | 172.69 | 3.12 |
| 15 | 5.42 | 0.9400 | -1.119 | 30.64 | 0.53 | 0.9860 | -1.019 | 171.27 | 3.89 |
| 19 | 5.14 | 0.8963 | -1.106 | 30.30 | 0.71 | 0.9860 | -1.007 | 169.10 | 3.88 |
| 33 | 16.92 | 0.9049 | -0.858 | 29.15 | 0.67 | 0.9902 | -1.013 | 171.54 | 3.25 |
| 5 | 7.70 | 0.9225 | -1.059 | 29.90 | 0.60 | 0.9916 | -1.033 | 172.81 | 3.00 |
| 9 | 11.23 | 0.9030 | -1.102 | 30.86 | 0.68 | 0.9914 | -1.028 | 172.84 | 3.03 |